%% file: sbgcinf_II_v11.tex
\documentclass[structabstract]{aa}  
\usepackage{mathptmx}
\usepackage{amssymb}
\usepackage{natbib}
\usepackage{graphicx,color}
\usepackage{txfonts}
\input{def}
\definecolor{lightred}{cmyk}{.1,1,1,0}
\definecolor{blueviolet}{cmyk}{.35,0,0.77,0.41}

\newcommand{\rem}[1]{}
\newcommand{\vsix}[1]{#1}
\newcommand{\vsev}[1]{#1}
\newcommand{\veight}[1]{#1}
\newcommand{\rev}[1]{{ #1}}
\def\CC#1{{  #1}}
\begin{document}
   \title{Superbubble dynamics in globular cluster infancy}
\titlerunning{Superbubbles in globular cluster infancy II}
   \subtitle{II. Consequences for secondary star formation 
     in the context of \\self-enrichment via fast rotating massive stars}

   \author{Martin Krause \inst{1,2,3}
     \fnmsep\thanks{E-mail: Martin.Krause@unige.ch}
          \and Corinne Charbonnel \inst{1,4}
          \and Thibaut Decressin \inst{1} 
         \and Georges Meynet \inst{1}
          \and Nikos Prantzos  \inst{5}
         %\and Roland Diehl \inst{2,1}
        }

   \institute{Geneva Observatory, University of Geneva, 
     51 Chemin des Maillettes, 1290 Versoix, Switzerland
     \and 
     Max-Planck-Institut f\"ur extraterrestrische Physik, 
     Postfach 1312, Giessenbachstr., 85741 Garching, Germany
     \and         
     Excellence Cluster Universe, Technische Universit\"at M\"unchen,
      Boltzmannstrasse 2, 85748 Garching, Germany
      \and
    IRAP, UMR 5277 CNRS and Universit\'e de Toulouse, 
     14 Av. E. Belin, 31400 Toulouse, France
     \and
     Institut d'Astrophysique de Paris, UMR7095 CNRS, 
     Univ. P. \& M. Curie, 98bis Bd. Arago, 75104 Paris, France
   }

   \date{Received: 5 November 2012; accepted: 5 February 2013}

% \abstract{}{}{}{}{} 
% 5 {} token are mandatory
 
  \abstract
  % context heading (optional)
   {\vsix{The self-enrichment scenario for globular clusters (GC) requires
       large amounts of residual gas after the initial formation of the first stellar
       generation. Recently, Krause et al. (2012) found that
       supernovae may not be
       able  to expel that gas, as required to explain their present day
       gas-free state, and suggested that} a sudden accretion on to
     the dark remnants, at a
     stage when type~II supernovae have ceased, may plausibly
     lead to fast gas expulsion. 
  }
% leave it empty if necessary  
  % aims heading (mandatory)
   {   Here, we explore the consequences of these results for the
     self-enrichment scenario via fast rotating massive stars (FRMS).}
  % methods heading (mandatory)
   { We analyse the interaction of FRMS with the
     intra-cluster medium (ICM), in particular where, when and how the
   second generation of stars may form. From the results, we develop a
 timeline of the first $\approx 40$~Myr {\CC{of GC evolution}}.}
  % results heading (mandatory)
   { The results of Paper~I imply three phases during which the ICM is
     in a fundamentally different state, namely the {{\em wind bubble
       phase}} 
      %CC ($\approx 3.5$ to {\CC{$8.8$}}~Myr), the {{\em supernova phase}} ($\approx$ {\CC{26.2 to}} $31.5$~Myr) and the {{\em dark remnant accretion phase}} ($\approx 0.1-4$~Myr):
        (\veight{lasting} $3.5$ to $8.8$~Myr), the {{\em supernova
            phase}} (\veight{lasting} $26.2$ to $31.5$~Myr), and the 
        {{\em dark remnant accretion phase}} (\veight{lasting} $0.1$ to $4$~Myr):
     (i) Quickly after the {\CC{first generation}} massive stars have formed, stellar 
     {{\em wind bubbles}} compress 
     the ICM into thin filaments. No stars may form in the
     normal way during this phase, due to the high Lyman-Werner flux
     density. 
     If the {\CC{first generation massive}} stars have however equatorial ejections, as {\CC{we}} proposed in
     the FRMS scenario, accretion may resume in the shadow of the 
     equatorial ejecta. The second generation stars may then
     form due to gravitational instability in these discs {\CC{that are fed by both the FRMS ejecta and pristine gas}}. 
    % The process may carry on through the 
     %{{\em supernova phase}}, because their ejecta, which observations
     %demand not to enrich the second generation stars, are not expected to
     %significantly interact with the discs. 
     (ii) \vsix{In the {{\em supernova
       phase}} the ICM develops strong turbulence, with characteristic
   velocities below the escape velocity.
   The gas does not accrete neither on to the stars nor on to
     the dark remnants in this phase due to the high gas velocities.
     The strong mass loss associated with the transformation of the
     FRMS into dark remnants then leads to the removal of the second
     generation stars from the immediate vicinity of the dark remnants.}
%      This means that secondary star formation may carry on undisturbedly for a
%      couple of Myr into the supernova phase, until also in the FRMS
%      with the lowest masses the equatorial ejection phase is over, and
%      all the second generation stars have been formed.
     (iii) When the
     supernovae have ceased, turbulence decays quickly, 
     and the gas can once more accrete, now
     on to the {{\em dark remnants}}.  As discussed in Paper I this may release sufficient energy to
     unbind the gas, and may happen
     fast enough so that a large fraction of less tightly
     bound first generation stars are lost.    
   }
  % conclusions heading (optional), leave it empty if necessary 
   {The detailed study of the FRMS scenario for the self-enrichment
     of GCs reveals the important role of the physics of the ICM in our
     understanding of the formation and early evolution of
     GCs. \rev{Depending on the level of mass segregation, t}his 
     sets constraints on the orbital properties of the stars, in
     particular high orbital eccentricities, which should have
     implications on the GC formation scenario.}
% The three ICM phases derived from the analysis of superbubble
%      dynamics in paper~I provide a - perhaps surprisingly - fitting
%      framework within which the
%      formation and enrichment of the
%      second generation stars may be understood.

   \maketitle
%
%________________________________________________________________

\defcitealias{Krausea12a}{Paper~I}

\section{Introduction}\label{intro}

Recent \CC{detailed} spectroscopic \CC{and deep photometric}  studies
have lead to the conclusion that  \CC{Galactic globular clusters (GCs) host 
multiple generations of stars;  those 
%CC can be distinguished from 
differ in their chemical abundance properties and in some cases in their 
%CC positions 
membership to multimodal sequences in the colour-magnitude diagram} \citep[e.g. ][ for recent
reviews]{GSC04,GCB12,Piotto09,Charb10}. 
The \CC{self-enrichment} scenario \CC{to explain the observations} is
likely complex. 
%CC \CC{In particular,}  \CC{the products of hot-hydrogen burning} are found in the second generation stars, \CC{while those of more advanced nuclear-burning phases} are not found \citep[compare also][paper~I hereafter]{Krausea12a}. 
%CC The \CC{self-enrichment} scenario \CC{to explain the observed abundance patterns} is likely complex. 
% In particular hot hydrogen burning products need to remain, supernova (SN) ejecta have to be removed.
\CC{It requires that the products of hot hydrogen-burning ejected by fast evolving first generation stars, remain within the GC and are recycled in second generation stars to explain e.g. the ubiquitous O-Na anti-correlation;
%CC and the multiple photometric sequences; 
but on the other hand, helium-burning products and supernova (SNe) ejecta have to be removed in order to keep the constancy of [C+N+O] and the mono-metallicity observed in most GCs. }

Two main models have been developed in the literature, according to the nature of the \CC{first generation stars potentially 
responsible for the GC self-enrichment (hereafter the stellar polluters), namely the fast-rotating-massive-stars (FRMS) or the asymptotic-giant-branch stars (AGB). 
Both the FRMS and the AGB scenarios face a problem in the mass budget between the amount of matter provided  by the polluter stars and the mass locked today in first and second generation low-mass stars \citep[about 30 vs 70~$\%$ respectively in most GCs, and up to 50-50 in few cases; for more details see ][]{Carrettaetal09}. 
Solutions require either a top-heavy IMF for the first stellar generation \citep{DantonaCaloi04,BekkiNorris06,PC06,Decrea07b}, or substantial loss of first generation low-mass stars from the GCs which must then have been much more massive initially than today  \citep{Decrea07b,Bekea07,DErcolea08,Decrea10,SC11}, the second option being currently favoured.}
In the FRMS scenario, \CC{the hydrogen-burning ashes} 
%CC second generation stars are polluted by slow equatorial winds, which 
would be ejected from their fast rotating parent stars due to
repeated super-critical rotation. \CC{These} ejecta are thought to mix with
pristine gas and then form the second generation
stars \CC{in the immediate surroundings of the massive star polluters.}  \citep{PC06,Decrea07b,Decrea07a}. 
Initial or very early mass segregation together with quick gas expulsion after the formation of the second stellar population, 
are invoked to loose preferentially first generation low-mass stars and
retain the
%CC  right mix 
observed proportion of first and second generation stars
\citep{Decrea10}.
In the asymptotic-giant-branch (AGB) scenario, the 
pristine gas left after the formation of the first stellar generation is supposed first to be expelled together with the
unwanted SNe ejecta and the \CC{bulk of} first generation low-mass stars. 
Later when the SNe would have ceased, 
the slow winds of AGB stars would accrete in a cooling flow towards the GC
centre, where they would form the second generation of stars \CC{after dilution with re-collected pristine gas whose sources are still uncertain }
\citep{DErcolea08,DErcolea10,DErcolea11,DErcolea12,CS11}\footnote{\CC{Dilution of the AGB ejecta with re-collected gas with pristine chemical composition is mandatory since all current AGB models \citep[e.g. ][]{KarakasLattanzio07,VenturaDAntona09,Siess10} predict that O and Na yields should actually correlate, at odds with the observed abundance anticorrelations. Although several scenarios have been proposed, the 
%CC sources and 
gathering mechanism of the required pristine matter \rev{is} still uncertain \citep[for discussion see ][]{DErcolea10,DErcolea11}.}}.
%Clearly, the gas expulsion by the SNe has so far been central for both scenarios.

%  The AGB scenario seems to fit nicely
% within the context of the general hydrodynamic evolution of spheroidal
% systems \citep[e.g.][]{DErcolea08,CS11}: 
% Here, the type~II supernovae (SNe) first clear out the cluster's
% gas. Then, there follows a phase of low energy input to the
% intra-cluster medium (ICM), coinciding with the slow AGB winds (30-50
% Myrs after the massive stars first form). SN~Ia are assumed to set
% in only after about 100~Myrs, and would again and finally clear out
% the gas in the GC. Thus, there is a time window when the AGB-ejecta
% would likely fall towards the centre and perhaps form stars. 

%-------------------------------------------------------------
   \begin{figure*}[!t]
     %\sidecaption
\centering
 \includegraphics[width=0.89\textwidth]{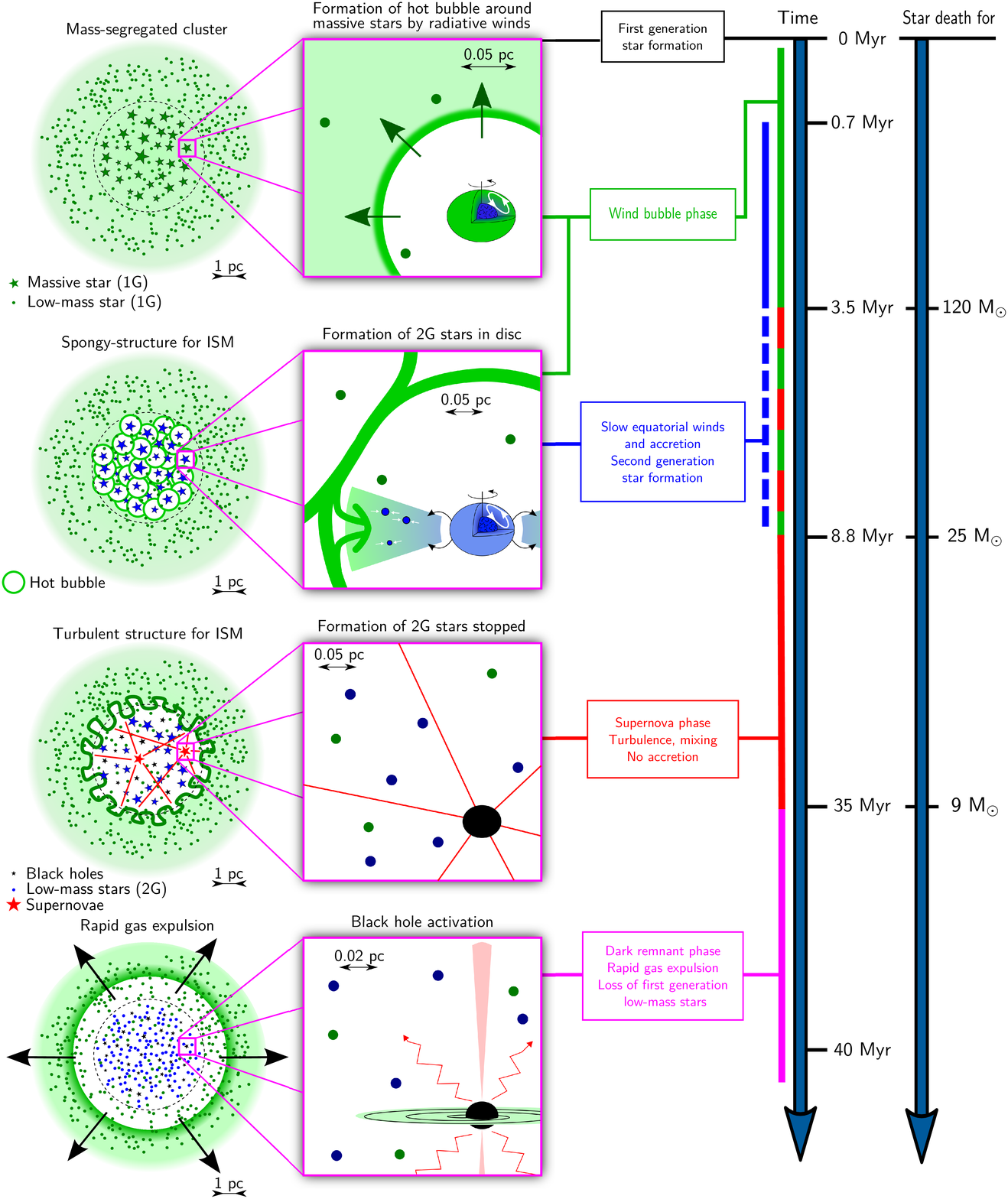}
%   \rotatebox{0}{\includegraphics[width=0.9\textwidth]{timeline3.eps}}
     \caption{\veight{Sketch of the proposed model of first 40~Myrs of evolution of a
two-populations globular cluster together with the global timeline
(top to bottom) and the associated time when the stellar life ends for
selected stars on the right. The leftmost column shows the whole
cluster. The second column from the left represents a zoom on to a
FRMS.  Four important stages are depicted.  \textit{First row from
top:} First a mass segregated star cluster is formed with 
\veight{all the FRMS inside the half-mass radius (dashed line)} 
and with the initial gas (light pistachio green shade) remaining after the star
formation.  Each massive star (the blue interior signifies the
convective hydrogen burning core, the bright forest green envelope
depicts the convective envelope, which has still the pristine
composition at this stage) creates a hot bubble around
it. The corresponding wind shell is shown in a darker shade of green
than the uncompressed gas. 
\textit{Second row:} All the hot bubbles connect and create a
spongy-structure in the centre. At the same time, slow mechanical
winds around the FRMS create a disk around them. The outer envelope of
the FRMS is now blue to denote that it has been contaminated by
hydrogen burning products. In the interaction
between the ejected disc (blue) and the accreting interstellar medium (ISM, green), 
a second generation of chemically different
stars is born (blue filled circles). 
\textit{Third row:} SNe (red stars with straight lines) fail to eject the gas but
create a highly turbulent convection zone. Further accretion on to the
remaining FRMS discs is inhibited. 
The equatorial ejections also end around this time and the
formation of the second generation stars is completed by about 10~Myrs
on the global time axis. \textit{Bottom row:} Later, rapid gas
expulsion takes place and removes all the remaining gas together with
the majority of the less tightly bound first generation stars out of
the cluster potential well. Such rapid gas expulsion is likely to be
due to the activation of black holes by accretion of matter.}}
         \label{fig:sketch}
   \end{figure*}
%-------------------------------------------------------------

In \vsev{\citet[Paper~I]{Krausea12a}} we showed that gas expulsion via SNe, which has \CC{long} 
been the prevailing paradigm \CC{to change the GC potential well and induce the loss of first generation low-mass stars}, does not work in
GCs. The reason is that, while the energy produced by SNe
usually exceeds the binding energy, 
%CC the energy 
it is not delivered fast enough to avoid the Rayleigh-Taylor instability of the escaping
shell. 
%This is a particular problem for early \CC{GCs}, as the gravitational
%attraction is strong on small \vsix{spatial} scales, and suddenly
%drops
%\vsix{around the core radius $r_\mathrm{c}$}, leading to
%strong acceleration.
We have also shown that a more powerful event, such as the energy
released by accretion on to the dark remnants, may lead to successful
gas expulsion.
These results suggest three distinct chronological phases in typical
GCs: 
\veight{Wind bubbles are created during the whole period from
  about~0 to 8.8 Myr on our global time scale.
  Core collapse to black holes and neutron stars occurs from about 
  3.5 Myr until 35 Myr. Where black holes are formed, the collapse may not 
  be accompanied by a strong or even any momentum ejection. If we
  therefore consider that only stars with initial masses below
  25~$M_\odot$ will give birth to an energetic supernova event then
  energy ejection by supernovae will occur only between 8.8 and 35
  Myrs on the global timescale.}
%In the first 3.5 \CC{to 8.8}~Myrs, the energy input is delivered by
%wind bubbles. Then type~II supernovae occur for about \CC{26 to}
%31.5~Myrs, \CC{with the actual duration of these first two phases 
%depending on the mass range for stars to directly form black holes 
%without exploding as SNe}. 
Finally, the dark remnants are activated and expel the gas 
\vsix{as well as the majority of the first generation stars}.

Here, we explore the implications of these findings for the formation
of the second generation stars in the context of the FRMS scenario and
develop a detailed timeline for the first $\approx40$~Myrs in the
lifetime of a typical GC. \veight{We describe the basic model setup in
  Sect.~\ref{sec:setup}. The wind bubble, supernova and dark remnant
  accretion phases are described in
  Sects.~\ref{sec:bubbles},~\ref{sec:SNe} and~\ref{sec:dmacc},
  respectively. 
%\rev{Important possible effects of a top-heavy IMF and
 % stronger mass segregation are explored in Sect.~\ref{sec:var}.} 
We discuss our findings in Sect.~\ref{sec:disc} and
  summarise and conclude in Sect.~\ref{sec:conc}.
In order to describe the sequence of
events, we refer to a global timescale throughout this article. The
global clock is set to zero at the coeval birth of the first
generation of stars.}

\section{Basic model setup}\label{sec:setup}
%\subsection{Model overview}
The model presented here generally follows the ideas
outlined in \citet{Decrea07a,Decrea10}.  We consider \vsix{an initial} dense gas cloud
of $M_\mathrm{tot}=9 \times10^6 M_\odot$ with a half-mass radius
of $r_{1/2}=3$~pc which forms \vsix{the first generation} stars at an efficiency of
$\epsilon_\mathrm{sf}=1/3$ and according to a Salpeter initial mass
function \rev{(IMF)} \CC{for first generation stars more massive than 0.8~M$_{\odot}$ and to a log-normal distribution for less massive long-lived stars of first and second generations}.
\CC{The parameters summarised in Table~\ref{table:pars} are inferred from N-body simulations which assume a Plummer distribution for the spatial distribution of gas and stars  (i.e., the star formation efficiency is similar at all \rev{radii}; \citealt{Decrea10})}. For our
model cluster, we find about 5700 \vsix{first generation} massive stars between 25 and
120~$M_\odot$. Mass segregation leads to all massive stars to be
located within $r_{1/2}$. We use the stellar evolutionary models of
FRMS at subsolar metallicity ($Z=0.0005$, \sqrbrk{Fe/H}$\eqsim-1.5$)
presented by \citet{Decrea07a}. From these models, we have extracted
the relevant wind parameters, summarised in
Table~\ref{tab:frms-windpars}. 
\citet{Decrea10} use these parameters to
describe the \rev{GC} \object{NGC~6752} with a present day
mass of $3\times10^5 M_\odot$, \CC{assuming complete recycling of the slow wind released by FRMS and dilution with pristine gas to reproduce the observed Li-Na anti correlation}.  For this setup, we will now explore
the different phases in detail. A sketch of the model is provided in
Fig.~\ref{fig:sketch}.

\begin{table}
\caption{Parameters of model cluster}             % title of Table
\label{table:pars}      % is used to refer this table in the text
\centering                          % used for centering table
\begin{tabular}{lrl}        % centered columns (4 columns)
\hline\hline                 % inserts double horizontal lines
Parameter & Value & Description\\    % table heading 
\hline                        % inserts single horizontal line
   $M_\mathrm{tot}$ & $9\times10^6M_\odot$ & Total mass\\     
   $r_{1/2} $ &3 pc& Half-mass radius\\
   $\epsilon_\mathrm{sf}$ & $1/3$ & Star formation efficiency \\
$\rho_0 $ & $10^6m_\mathrm{p}$~cm$^{-3}$ & Average gas density \\
\hline                                   %inserts single line
\end{tabular}
\end{table}
%
%_____________________________________________________________

\section{Hot wind bubbles}\label{sec:bubbles}
We will now argue (Sect.~\ref{sec:bubexp})
that during the first few Myrs of the GC, before the
first type~II SNe, the fast radiative winds of the massive stars will create
large bubbles that will probably also overlap, but they will probably
not unite into a single superbubble 
\veight{and will not lift any important amount of gas out of the GC.} 
This situation is depicted in Fig.~\ref{fig:sketch}
(second row from top). \veight{We also show that during this wind
  bubble phase (and also the supernova phase), no
stars should form in the normal way (Sect.~\ref{sec:lwflux}), the
equatorial ejections of the FRMS establish a decretion disc close to
the FRMS (Sect.~\ref{sec:eqej}) and that accretion may be
re-established in the shadow of these decretion discs
(Sect.~\ref{sec:acc}). The inner decretion and the outer 
accretion discs merge on the viscous timescale
(Sect.~\ref{sec:discmerge}), and finally the second generation stars
form in the merged discs (Sect.~\ref{sec:sfd}).}

\subsection{Bubble expansion}\label{sec:bubexp}
To see that the wind bubbles may not unite into a single superbubble
and that they will also not lift much gas out of the GC's potential
well, we consider the energy budget in the cluster.
\veight{
\citet{FHY03} derive that the mechanical power output of a star
dominates the effect on the intra-cluster medium (ICM) rather than the
effect of ionisation.}
The power output of the
fast winds of a FRMS with a given mass $M$ 
\vsix{between $40 M_\odot$ and $120 M_\odot$} may be approximated by
\eq{Q_0(M) = 4 \times 10^{30} (M/M_\odot)^{2.8} \;\mathrm{erg\;s}^{-1}.} 
This power law
follows from interpolation of the values in
Table~\ref{tab:frms-windpars}. 
With a Salpeter \rev{IMF}, the total wind output, which is dominated by
the most massive stars, integrates to:
\rev{$Q_\mathrm{0,tot}=2\times10^{39}$~erg~s$^{-1}=6\times10^{52}$~erg~Myr$^{-1}$.} 

How much of this energy is actually used to extend the bubbles and not
lost to radiation?
\citet{FHY03} derive that for a background density of $20$~cm$^{-3}$,
\veight{much lower than the density expected in a young \rev{GC},
about 90 per cent of this power is radiated away, and thus the energy
efficiency is only 10\%. If the bubble would
remain self-similar, a 
greater energy efficiency is expected, around 70\%
%smaller fraction of about 3/11 of the
%total power would be expected to be radiated 
\citep{Weavea77,FHY03}.
\citet{Krausea13a}
study the energy efficiency of merging wind bubbles, and find that for
small separations, there is no additional effect from bubble merging
and the efficiency factor is not changed.} We expect that in the
dense environment of a forming GC radiation losses may
increase and the efficiency of
energy transfer may be reduced below the value of \citet{FHY03}. On the
other hand, the optical depth is higher, and more of the radiation
energy may be captured by the gas. In the following, we adopt
$\eta=0.1$ as a working hypothesis for the energy efficiency,
but the conclusions would be unchanged if $\eta=0.01$.

The gravitational binding energy of the gas is
\vsix{$E_\mathrm{grav}=0.4(1-\epsilon_\mathrm{sf}) GM^2/R 
\approx 6 \times10^{53}$~erg \citep{BCP08}}. 
The combined wind power of
all the massive stars amounts \rev{to $\eta
Q_\mathrm{0,tot}=6\times10^{51}(\eta/0.1)$~erg~Myr$^{-1}$.} Thus, it is not
possible for the stellar winds to lift any noteworthy amount of gas
out of the GC on a relevant timescale.

\begin{figure}
   \centering
   \includegraphics[width=.5\textwidth]{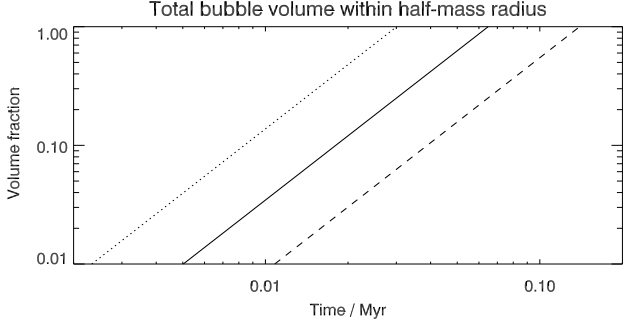}
      \caption{Combined volume of the wind bubbles of all the 5700
        massive stars ($M>25M_\odot$) in our model cluster as fraction
        of the half mass radius. The solid (dotted, dashed) line is for an energy
        injection efficiency of $\eta=0.1$ (1, 0.01).         
              }
         \label{fig:bubvol}
   \end{figure}

On the other hand, we may integrate the volume fraction occupied by
the wind bubbles by integration of the standard wind bubble volume
$4\pi r^3_\mathrm{bubble}/3$ over the IMF. The bubble radius
$r_\mathrm{bubble}$ at time $t$ is given by:
\eql{rbub}{r_\mathrm{bubble}=\alpha
  \left(\frac{\eta Q_\mathrm{0,tot}}{\rho_0}\right)^{1/5}
  t^{3/5}\, ,}
where $\alpha\approx0.8$ for self-similar expansion
\citep{Weavea77}, and the not necessarily self-similar thin shell
approximation alike \citep{mypap03a}. The average gas density in our model
cluster is $\rho_0=10^6m_\mathrm{p}$~cm$^{-3}$. We include the
efficiency factor $\eta$ in \rev{Eq.}~\ref{rbub}, because the expansion
law in this form is consistent with the sizes of observed wind bubbles
\citep[e.g.][ and references therein]{OG04}. 
We show the volume fraction occupied by hot bubbles in
Fig.~\ref{fig:bubvol}. The hot wind bubbles fill the GC very quickly
on a timescale of~0.1~Myr. 
%We note that this timescale is only
%insignificantly prolonged if one considers the additional energy
%needed to lift up some of this gas (e.g. constant density or density
%growing with $1/r^2$)
%from the immediate vicinity of the massive stars.

%-------------------------------------------------------------
%-------------------------------------------------------------

Thus, while the wind bubbles are not able to lift the gas out of the
potential well of the GC, they are nevertheless able to quickly fill the entire volume
within the half-mass radius, no matter what we assume in detail for
the energy efficiency. It follows that the bulk of the cluster gas is
compressed into thin filaments or sheets, while most of the volume is
occupied by hot wind bubbles. Bubbles of stars of different masses in
general do not have the same pressure, and thus many individual
interfaces may be expected to be swept away and pushed against some
other bubble walls \citep{vanMea12,Krausea13a}. 
So, overall it is clear that some individual
bubbles will unite to form larger bubbles, but there will not be one
overall superbubble which would imply significant lift-up of gas out
of the potential well of the GC.

%%%%%%%%%%%%%%%%%%%%%%%%%%%%%%%%%%%%%%%%%%%%%%%%%%%%%%%%
%\setcounter{table}{0}
\begin{table*}[t]
\caption{\label{tab:frms-windpars} Properties of low metallicity fast rotating massive stars.}
\centering
\begin{tabular}{@{}cccccccccc@{}}
    \hline\hline
     $M_*$\tablefootmark{a} / & 
     $R_*$\tablefootmark{b} / &
     $\dot{M}_\mathrm{fast}$\tablefootmark{c} /  &
     $\dot{M}_\mathrm{slow}$\tablefootmark{d} / & 
     $T_\mathrm{eq,start}$\tablefootmark{e} / &
     $T_\mathrm{eq,end}$\tablefootmark{f} / &
      ${\Delta}t_\mathrm{on}$\tablefootmark{g} / 
     & ${\Delta}t_\mathrm{off}$\tablefootmark{h}/
     & $Q_{0}$\tablefootmark{i}     / 
     & $Q_\mathrm{LW}$\tablefootmark{j}     / \\

    $M_\odot$ &$R_\odot$&$M_\odot$~Myr$^{-1}$&$M_\odot$~Myr$^{-1}$ & 
     Myr&Myr&Myr&Myr&$10^{35}$erg~s$^{-1}$&$10^{49}$s$^{-1}$\\

     \hline

     120 & 12& 1     & 66 & 0.7 & 3.15 & 0.020 & 0.080 & 28 & 9\\ 
     % times 10^47 phot/s/(all stars) Ly-We-flux
     60  & 10& 0.2   & 14 & 0.9 & 4.4 & 0.028 & 0.067 &   3.9 & 1.26\\ %so, we confirm 
     40  & 8& 0.06 & 14 & 1.1 & 5.7 & 0.065 & 0.250 & 1.2 & 0.192\\ %Conroy & Spergel 2011
\hline
\end{tabular}
\tablefoot{ 
\tablefoottext{a}{Mass of the star.}
\tablefoottext{b}{Photospheric radius.}
\tablefoottext{c}{Mass loss rate of the steady fast wind.}
\tablefoottext{d}{Mass loss rate of the slow equatorial wind.}
\tablefoottext{e}{Global time at which the intermittent equatorial ejections start.}
\tablefoottext{f}{Global time at which the intermittent equatorial ejections end.}
\tablefoottext{g}{Duration of the individual slow wind episodes.}
\tablefoottext{h}{Time span in between any two slow wind episodes.}
\tablefoottext{i}{Total power of the steady, radiatively driven fast
  wind on the main sequence.}
\tablefoottext{j}{Flux of Lyman-Werner photons ($912\AA$ -- $1100\AA$)
  at 1~Myr after reaching the main sequence; roughly constant during
  the main sequence.}
}
\end{table*}

\subsection{Lyman-Werner flux}\label{sec:lwflux}
\citet{CS11} make the point that during the main sequence phase of the
massive stars, the Lyman-Werner flux ($912\AA$ -- $1100\AA$) is so
high that molecular hydrogen is dissociated throughout the cluster,
and no stars may form in the usual way during this time. We repeat
their analysis here, but with the numbers from the FRMS models of
\citet{Decrea07a}. We have calculated the Lyman-Werner flux from these
models and included it in table~\ref{tab:frms-windpars}. A
reasonable fit to the photon fluxes, which are roughly constant during
the main sequence, as a function of mass of the
parent star is:
\eq{Q_\mathrm{LW}=7\times10^{43} (M/M\odot)^{2.9} \;\mathrm{s}^{-1}\;
  .}
\citet{CS11} estimate the radius of the photo-dissociation region to
\eq{R_\mathrm{LW}=1.1\;\mathrm{pc}\;
\left(\frac{Q_\mathrm{LW}}{10^{49}\;\mathrm{s}^{-1}}\right)^{1/3}
\left(\frac{10^6 m_\mathrm{p}\mathrm{cm}^{-3}}{\rho}\right)^{1/3}
\left(\frac{0.028 Z_\odot}{Z}\right)^{1/3}
\; ,}
adapted to the metallicity of our model stars.
Thus, a $120 M_\odot$ star is already able to
photo-dissociate a large fraction of our model GC. 
Integration over a Salpeter IMF gives the factor $f_\mathrm{LW}$,
which describes how many times the GC volume could be photodissociated by the
ultraviolet radiation:
\eq{f_\mathrm{LW}=200
\left(\frac{10^6 m_\mathrm{p}\mathrm{cm}^{-3}}{\rho}\right)
\left(\frac{0.028 Z_\odot}{Z}\right)\, .}
So, even if the metallicity or the gas density would be higher by one
or even two orders of magnitude, the gas would still be fully 
photodissociated. 

\rev{In addition to the effects of dust considered in the above
  analysis, molecular hydrogen may also form via electrons, which
  first combine with neutral hydrogen to H$^-$. H$_2$ is then formed
  via the reaction:
\eq{\mathrm{H}+\mathrm{H}^- \rightarrow \mathrm{H}_2 + e \, .}
Regions where both, electrons and neutral hydrogen atoms, are abundant
are rare. \citet{RGS01} show that the 
the interface between an H~{II\sc} region and a neutral region is such
a promising place, and that
molecular hydrogen may form there in a narrow layer. \citet{RGS01}
find a strong dependence on the input spectrum and a maximum column density of
$\log(N_{\mathrm{H}_2}/\mathrm{cm}^{-2})=14-15$ of molecular hydrogen
formed in this layer. The corresponding gas mass itself would be too
small to form any significant amount of stars in the GC
context. However, \citet{RGS01} also show that the column is sufficient for
some Lyman-Werner bands to become optically thick. Hence, such layers
might in principal shield pockets of dense gas from
photo-dissociation. Quantitatively,  \citet{RGS01} find that the
radius of photo-dissociation regions may be reduced by a factor of~1.5,
corresponding to a factor of~3.4 in volume. This would however be too
small to be significant, since we have shown above that there are
enough photons to ionise the entire volume about 200 times. This crude
order of magnitude estimate stands up to recent three-dimensional
radiative transfer simulations: \citet{WHB11} show
that passage through molecular hydrogen with a column density of 
$N_{\mathrm{H}_2} \equiv  5x \times 10^{14}\mathrm{cm}^{-2}$
attenuates the Lyman-Werner flux (within an accuracy of 15~per~cent) 
by factor 
\eqs{f_\mathrm{shield}(N_{\mathrm{H}_2},T) &=& \nonumber
\frac{0.965}{(1+x/b_5)^{1.1}} + \frac{0.035}{(1+x)^{0.5}} \\
& &\times \exp \left[ -8.5\times 10^{-4} (1+x)^{0.5}\right]\, ,}
where $b_5\equiv  b /(\mathrm{km}\, \mathrm{s}^{-1})$ is the 
Doppler broadening parameter.
If thermal broadening dominates, $b_5$ should be around or below
unity. If their is substantial velocity shear in the filaments, it
might be higher. A reduction of the Lyman-Werner flux by a factor of
200 (in order to get the photo-dissociated volume fraction below
unity) 
is reached  at $\log(N_{\mathrm{H}_2}/\mathrm{cm}^{-2})=17$
(16,18) for $\log(b_5/\mathrm{km}\, \mathrm{s}^{-1})=0$ (-1,1).
This confirms
that the aforementioned layers at the interface of ionised and neutral
regions with $\log(N_{\mathrm{H}_2}/\mathrm{cm}^{-2})=14-15$ would not 
provide a significant optical depth which
could significantly alter the conclusion above. A combination of this
effect together with strong dust absorption might perhaps allow for some
molecular hydrogen in the densest parts of the ICM. However, the
question would remain, why these dense pockets would not have formed
stars right away.
 }

We therefore confirm the result of
\citet{CS11} with the star models of \citet{Decrea07a} 
that molecular hydrogen is no longer present \rev{in significant
  amounts} in the GC
as soon as the massive stars reach the main sequence.
From the similarity of the numbers for our star models at a metallicity of
$Z=0.03Z_\odot$, with their case for $Z=0.1Z_\odot$, it
should follow that this result should not depend very much on the
exact metallicity of the GC.
Thus, at least low-mass stars do not form 
\vsix{in the ``classical'' way} during this time, because
the gas is efficiently kept at $\approx100$~K.
\veight{We note that this remains true throughout both,
the wind bubble and the supernova phase.}

\subsection{Equatorial mass ejections}\label{sec:eqej}
Angular momentum transfer within the FRMS causes episodic equatorial
mass ejections of gas enriched by hydrogen-burning products 
when the star reaches break-up,
forming decretion discs \citep{Decrea07a,KOM11} around
the stars.
\vsix{This process is analogous to the equatorial mass loss in Be-type
stars \citep[e. g.][]{Riviea01,TOH04,Haubea12}.
In FRMS, the total amount of} mass lost by this process is rather high, and it is hardly
plausible that the ejected material may immediately leave the vicinity
of the star: In order to lift the material out of the gravitational
potential of the parent star, an energy supply at a rate of $GM
\dot{M}/(2R_\mathrm{dd})$ is required, if the decretion disk is first
established at a characteristic radius $R_\mathrm{dd}$ after being
ejected. The factor of two in the denominator takes into account the
rotational energy of the gas. This material leaves the star because of supercritical
rotation. So, it is reasonable to assume that it will be ejected with the
necessary angular momentum to be stable just outside the star. As an
estimate, we take $R_\mathrm{dd}$ to be three photospheric radii 
(compare Table~\ref{tab:frms-windpars}).
For the 40 (60,120) $M_\odot$ star, we find a required energy injection rate of
1.4~(1.7,13)$\times10^{36}$~erg/s. 
The power in the fast, radiative wind is a few, up to a
factor of ten times 
less\footnote{Our wind power $Q_0$ refers to the terminal wind power
  of the fast, radiatively driven wind,
  but the additional power required to lift the material ejected in
  this wind out of the gravitational potential is less than 50~per cent
of $Q_0$.}. 
%Radiation pressure acts more efficiently on gas with low
%optical depth. Additionally, t
The equatorial ejections should cover
only a small part of the solid angle \citep{KOM11}, whereas at the
same time the radiation of fast rotating stars is concentrated towards
the poles \vsix{\citep[e.g.][]{MM12}.} In the light of these
considerations, it seems highly
implausible that a significant part of the disc may leave the
immediate vicinity of the parent star, though, there may likely be
some moderate mass loss via photo-evaporation, and the disc may 
spread out somewhat via angular momentum transfer \citep{KOM11}.

The mass lost by the equatorial mechanism averaged over \vsix{the total
ejection time, which includes the main sequence and the luminous blue
variable stages,} may be well fitted by the formula:
\eql{eq:eqej}{
\dot{M}_\mathrm{eq}=
 \left[9 \left(\frac{M}{60M_\odot}\right) -4.8 \right] M_\odot
 /\;\mathrm{Myr}\, ,}
for stars with mass $M>32 M_\odot$. \vsix{The latter limit is simply due to the
zero-point in \rev{Eq.}~(\ref{eq:eqej}). This is however quite realistic,
as the 40~$M_\odot$ model still has a high equatorial mass loss rate,
whereas in the 25~$M_\odot$ model it is close to zero.}
%Once, all stars are active
%($\approx 1$~Myr after birth of the population), the total mass output
%is of order $10^4 M_\odot/$Myr.

\subsection{Accretion on to the discs around the massive stars}
\label{sec:acc}
\subsubsection{The discs persist around the FRMS}
The off-time between any two equatorial mass ejections is a few times
the on-time (Table~\ref{tab:frms-windpars}). Also the radiative energy
accumulated in the off-time, when no new mass is ejected equatorially, but the
radiation pressure keeps pushing, is likely not sufficient to unbind
the entire disc from the parent star.
Another process to consider is close encounters of other massive stars, which
could in principle unbind a disc \citep{CP93,HCP96}. In the
unfavourable prograde encounter, the disc would be stripped down to
about half the periastron radius. The timescale for a given massive
star to have a close encounter with another massive star is given by
$(n \sigma v)^{-1}$, where $n= 20,000/(4 \pi
r_{1/2}^3/3)=192$~pc$^{-3}$ is the density of massive stars, 
$\sigma=\pi (2 R_\mathrm{dd})^2$ is the characteristic cross section
of the accretion disc, and $v\approx 100$~km/s is the typical velocity
of the massive stars.  Using any of the values for the disc radii
discussed above, the encounter timescale turns out to be much larger than
the Hubble time. Thus, the discs cannot be tidally stripped.
\veight{We note that, due to the small cross section of the FRMS
  discs, ram-pressure stripping is equally ineffective.}

\subsubsection{Re-establishment of accretion}
Importantly, the shielding from the radiation
of the parent star should allow a re-establishment of the accretion
flow providing pristine gas (Fig.~\ref{fig:sketch}, \CC{second row from top, right}). 
The Bondi accretion rate should set an upper limit for
the mass accumulation in the disc.
%, which would be reached, if the
%accreting matter would have no angular momentum. 
With the sound speed
for an atomic gas consisting of 90~per cent hydrogen and 10~per cent
helium by number, $c_\mathrm{s}^2= \gamma k_\mathrm{B} T / (1.27
m_\mathrm{p})$, and a solid angle fraction $f_\mathrm{sd}$ of the disc
usable for accretion, the Bondi accretion rate evaluates to \citep{FKR02}:
\eqsl{eq:Bondi}{\dot{M}_\mathrm{Bondi}&=&4600 M_\odot /\;\mathrm{Myr}\nonumber\\
& &\times\, \left(\frac{f_\mathrm{sd}}{0.1}\right)
\left(\frac{M}{60M_\odot}\right)^2
\left(\frac{\rho}{10^6m_\mathrm{p} \mathrm{cm}^{-3}}\right)
\left(\frac{100\,\mathrm{K}}{T}\right)^{3/2}\, .
}

%CC This accretion rate is most likely not sustained throughout the lifetime of the massive stars. One reason is that the dense gas should be compressed in filaments at this time (compare above), so that accretion may only take place when the star passes through a filament. Yet, since the dependence on the density is linear, we do not need to take this into account for the present discussion, as a reduced accretion rate in voids would be compensated by an increased one in the filaments. If a star moves with a velocity $v$ relative to the local ICM, the accretion rate is reduced by a factor $(v/c_\mathrm{s})^3$ \citep{BH44}. 
\vsix{This accretion rate is most likely not sustained throughout the
lifetime of the massive stars. One reason is that the dense gas should
be compressed in filaments at this time (compare above), so that
accretion may only take place when the star passes through a
filament. We show below that, for suitable orbit parameters, 
the total path an FRMS traverses at low
velocities is long compared to the typical bubble size.
Since the dependence on the density is linear, we therefore do
not need to take this into account for the present discussion, as a
reduced accretion rate in voids would be compensated by an increased
one in the filaments.} 

\subsubsection{Suppression of accretion due to stellar velocities}
\label{subsec:accsup}
If a star moves with a velocity $v$ relative to the local
ICM, the accretion rate is reduced by a factor
$(v/c_\mathrm{s})^3$ \citep{BH44}.  
For typical virial velocities of
about 100~km/s, this would lead to a suppression of about six orders of
magnitude. Orbits of stars in \rev{GC}s are not \vsix{expected
to be} circular, and
therefore the accretion rate is a strong function of the
position of a given star along its orbit: It
accretes more strongly when the velocity relative to the ICM is small, i.e.
near the outer turning point and when at the same time it passes
through a filament. 

%-------------------------------------------------------------
   \begin{figure}
   \centering
   \includegraphics[width=.5\textwidth]{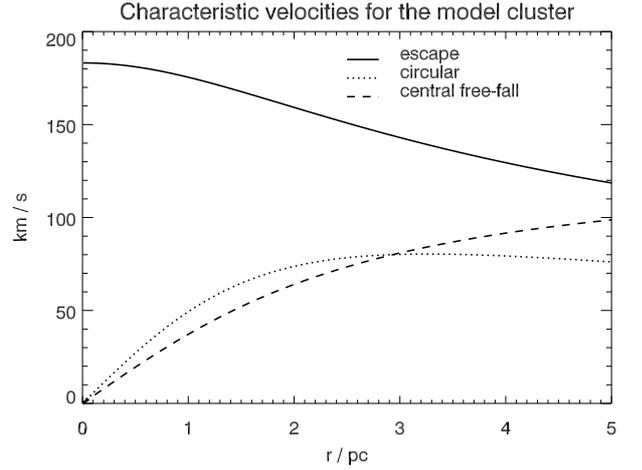}
      \caption{Characteristic velocities for the model GC 
        (compare Table~\ref{table:pars}) as a function of radius. The
        solid line shows the local escape velocity. The dotted line is
        for the circular velocity. The dashed line shows the velocity
        a star would have obtained in the centre if it would start at
        rest at a given radius $r$.      
              }
         \label{fig:vels}
   \end{figure}
%-------------------------------------------------------------
%-------------------------------------------------------------
   \begin{figure}
   \centering
   \includegraphics[width=.5\textwidth]{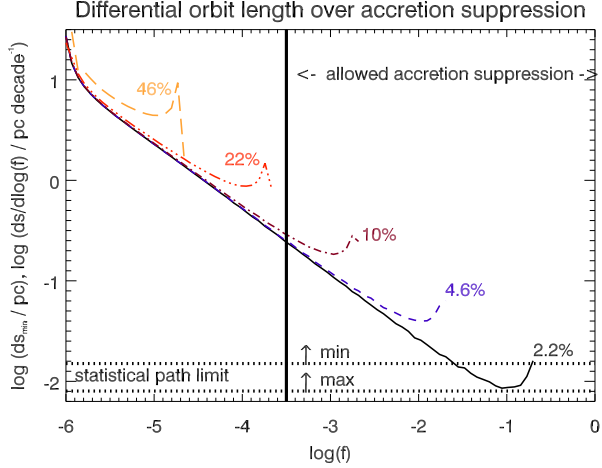}
      \caption{Differential path length over accretion suppression
        factor $f=<{\dot{M}}_\mathrm{Bondi}>/{\dot{M}}_\mathrm{Bondi}$. The five
      curves labelled by their respective
~      $v_\mathrm{az,0}/v_\mathrm{circ} $-value show the path length in
      parsec that a star spends at a given accretion suppression factor
      $f$ per orbital period and per decade in log$(f)$. The range of 
      $f$-values that our model requires is to the right of the solid
      vertical bar. \veight{The horizontal dotted lines show the limiting path 
      length for statistically uniform accretion. 
      The plot shows that the accumulated path length is sufficient
      for uniform accretion for all orbits with
      $v_\mathrm{az,0}/v_\mathrm{circ} \gtrsim 3\%$.  For orbits with
      smaller values of $v_\mathrm{az,0}/v_\mathrm{circ}$, accretion
      varies strongly from star to star, and the average
      value shown in Fig.~\ref{fig:sup} would only apply to the
      ensemble. The plot also shows, that for all orbits with
      $v_\mathrm{az,0}/v_\mathrm{circ} < 10\%$ there is always
      a basic amount of gas that is uniformly accreted on to each FRMS
      disc. This amount is sufficient for our model. The part of the 
      accreted gas that varies from star to star comes only on top of
      this basic amount.
%But, because the differential path length curves
%     reach higher $f$-values in this case, the amount of accreted gas
%     for each encounter between a star at low velocity and a filament
%    is increased.
%per orbit for a given star to
 %     accumulate a total path of one typical bubble diameter over the
  %    minimum (upper line) and maximum (lower line) total accretion
  %    timescale. 
      See section~\ref{subsec:accsup} for details.}
              }
         \label{fig:path}
   \end{figure}
%-------------------------------------------------------------
%-------------------------------------------------------------
   \begin{figure}
   \centering
   \includegraphics[width=.5\textwidth]{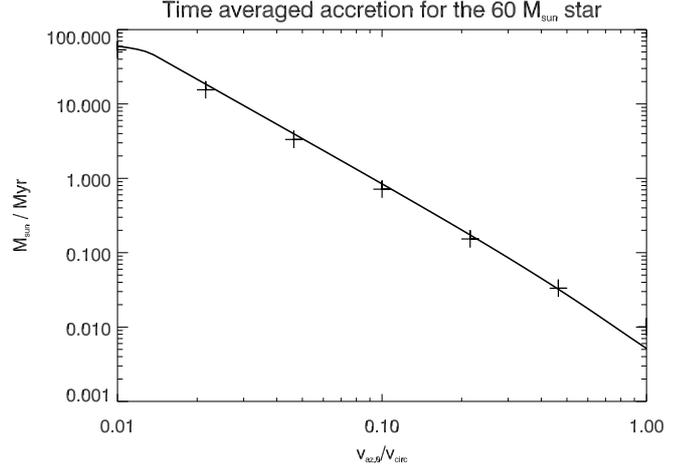}
      \caption{Dependence of the time averaged accretion rate on to
        the FRMS as a function of the \vsix{ratio of the azimuthal velocity at the outer
        turning point,
        $v_\mathrm{az,0}$, over the circular velocity
        $v_\mathrm{circ},$ for the $60 M_\odot$ star}
        (see section~\ref{sec:acc} for details). The solid line shows the elliptical
        approximation. The pluses show the result of a direct orbit
        integration in the actual Plummer potential. There is
        excellent agreement between the two methods.      
              }
         \label{fig:sup}
   \end{figure}
%-------------------------------------------------------------
 We will now estimate the time averaged accretion rate for a given
star.
In a spherical potential, the orbits are generally
expected to be of the rosette type \citep{BT08}, with most of the
stars having only slightly negative total energies. Adapted to our
model cluster, most of the massive stars will have just enough energy
to reach the core radius at their outer turning points. 
As may be seen from Fig.~\ref{fig:vels}, the circular velocities at
the core radius (2.3~pc) are
around 80 km~s$^{-1}$. If the stars would have no angular
momentum, they  would reach a similar velocity after a free fall to
the centre from this position. Consequently, velocities around 1~km/s,
as necessary for efficient accretion may only be reached for very
eccentric orbits and only near their outer turning points.

For this estimate, we approximate the orbits by ellipses.
The radius $r$ as a
function of azimuth $\phi$ is then given by 
\eq{r=\frac{p}{1+e \cos(\phi)}\, ,}
with the eccentricity $e$ and the orbital size parameter $p$.
The orbital velocity follows by differentiation and may be expressed
as
\eq{v_e(\phi)=v_0 \sqrt{\frac{1+2e\cos \phi + e^2}{1-e^2}}\, ,}
where $v_0^2 = G M_\mathrm{tot}/a$ with the semi-major axis $a$. 
\veight{Instead of the eccentricity, we use in the following the
ratio of the velocity at the outer turning point to the circular
velocity at this location as parameter. It is given by 
$v_\mathrm{az,0}/v_\mathrm{circ}=\sqrt{(1-e)/(1+e)}v_0/v_\mathrm{circ}$.}
The orbit-averaged Bondi accretion rate may then be found by integration
of \rev{Eq.}~(\ref{eq:Bondi}) over the appropriate number of orbits,
$n$, multiplied by the suppression factor, 
\veight{$(c_\mathrm{s}/v_e)^3$,} and dividing by $n$ times the orbital period $T$:
\eql{eq:mdotav}{<{\dot{M}}_\mathrm{Bondi}> =\dot{M}_\mathrm{Bondi}
  \frac{1}{nT}\int_0^{nT}\left[\mathrm{min}\left(1,\frac{c_\mathrm{s}}{v_e}\right)\right]^3 
\mathrm{d}t} 
%I(e) \\
%I(e) &=& 
%\int_0^{2\pi} \frac{\mathrm{d}\phi}
%{(1-2e\cos\phi+e^2)^{3/2} (1+e\cos\phi)^2}\,.}
\veight{We remind here that ${{\dot{M}}_\mathrm{Bondi}}$ depends on density
and temperature. We have taken this out of the integral in 
\rev{Eq.}~(\ref{eq:mdotav}), because the
changes in density along the path of a given star --
due to filaments and voids -- are not correlated to the changes in
velocity. Again, since the Bondi accretion rate is linear in the
density, we may use the average value for the density as long as we
make sure that the total path traversed while the velocity of the star
is small (the contribution to the integral is negligible otherwise) 
is long compared to the average bubble diameter. This is
indeed the case for an interesting part of the parameter space in 
$v_\mathrm{az,0}/v_\mathrm{circ}$:}
We parametrise the accretion suppression by
$f=<{\dot{M}}_\mathrm{Bondi}>/{\dot{M}}_\mathrm{Bondi}$
and show the path-length per orbit per decade of log($f$) in
Fig.~\ref{fig:path}. The orbital period for stars which have the
outer turning point near the core radius is
0.16~Myr for all eccentricities. The minimum
accretion time is given by the equatorial ejection phase of the
120~$M_\odot$ star, 2.45~Myr. The maximum accretion time would be
4.6~Myr, which would correspond to the 40~$M_\odot$ star and the case
when the massive stars turn into black holes without energetic
SNe. Hence, the FRMS complete 15 to 29 orbits in the GC 
\veight{during the
time when they have equatorial ejections.} 
In order to accumulate a total path at low velocity of a
bubble diameter, they must therefore spend at least 0.008~pc or,
respectively, 
0.015~pc at low velocities, per orbit. These two limits
are shown as dotted horizontal lines in Fig.~\ref{fig:path}. A
vertical solid line marks the minimum accretion rate we must demand to
explain the relative contributions of pristine and ejected gas in the 
presently observed GC stars. \veight{It can be seen that, for small values of
$v_\mathrm{az,0}/v_\mathrm{circ}$, the differential path
length distribution is well described by a power law, 
\dm{\frac{\mathrm{d}s}{\mathrm{dlog}(f)}\propto f^{-3/5}\, ,}
with an upturn at high values of log$(f)$. Figure~\ref{fig:path} shows
that the accumulated path is generally above the statistical limit for
$v_\mathrm{az,0}/v_\mathrm{circ} \gtrsim 3\%$. For such orbits, 
\rev{Eq.}~\ref{eq:mdotav} is therefore valid for each star. Below this
value, \rev{Eq.}~\ref{eq:mdotav} is only valid for an ensemble that is
sufficiently large. Figure~\ref{fig:path} also shows that for all
stars with $v_\mathrm{az,0}/v_\mathrm{circ} < 10\%$, the integrated
path between log$(f)=-3.5$ and log$(f)=-2.5$ is always much larger
than the statistical path length limit. Therefore, such orbits have a
uniform basic accretion rate, which is sufficient in the context of
our model. Only on top of this, there can be a strongly varying component.
Within these limitations, we may therefore use
\rev{Eq.}~\ref{eq:mdotav} to describe the average accretion rate.}

The orbit-averaged accretion rate for
the 60~$M_\odot$ star as a function of
$v_\mathrm{az,0}/v_\mathrm{circ} $
is shown in Fig.~\ref{fig:sup}. We have also
derived the orbit-averaged accretion rate by direct numerical
integration of the orbit in the Plummer potential, which is shown as
pluses in the same plot. The good agreement is expected, as the 
inner regions, where \vsev{the} elliptical approximation is bad, do not
contribute much to the average accretion rate. 
\veight{We note that the time averaged accretion rate may be well
  approximated by:
\eqsl{eq:mdotfit}{\dot{M}_\mathrm{Bondi}&=&0.7 
  \left(\frac{v_\mathrm{az,0}/v_\mathrm{circ}}{0.1}\right)^{-2} M_\odot
  /\;\mathrm{Myr}\nonumber\\
& &\times\, \left(\frac{f_\mathrm{sd}}{0.1}\right)
\left(\frac{M}{60M_\odot}\right)^2
\left(\frac{\rho}{10^6m_\mathrm{p} \mathrm{cm}^{-3}}\right)
\left(\frac{100\,\mathrm{K}}{T}\right)^{3/2} \, ,
}
which is fitted to the curve in Fig.~\ref{fig:sup}.}

While the
characteristic velocity $v_0$ leads to a suppression of the accretion rate
by about six orders of magnitude, high orbit eccentricity may 
compensate this to some degree.
 An accretion rate of about 1.5~$M_\odot$/Myr, as required to get
the right \CC{dilution (i.e., that requested to explain the observed Li-Na anticorrelation)} of ejected and pristine gas (compare \rev{Eq.}
(\ref{eq:eqej})), is reached for $v_\mathrm{az,0}/v_\mathrm{circ}= 7\%$, for the 60~$M_\odot$ star. 
In this case, the gas in the disc around
the  60~$M_\odot$ star would consist -- on average -- of 74~\% ejected
gas from the parent star, and 26~\% of accreted pristine gas. The
accretion rate grows quadratically with the mass of the accreting star, 
whereas for the equatorial mass ejection rate the dependency is
linear. The pristine gas fraction varies therefore with the
mass of the parent star. At 60~$M_\odot$ it is close to minimal, for
40~(120)~$M_\odot$ it rises to 35~\% (31~\%), respectively, assuming
again 
$v_\mathrm{az,0}/v_\mathrm{circ} = 7\%$,
We discuss this further in section~\ref{subsec:pri-to-pro}.

\subsection{Viscous merging of accretion and decretion discs}
\label{sec:discmerge}
Because of the angular momentum that is likely present in the gas, it
will settle into the disc at some radius $r$, which is likely large
compared to the typical locus of the equatorial stellar ejecta.
Viscous processes are then responsible for transporting the material
within the \vsev{accretion} disc. 
For a steady, thin, Keplerian disc, the viscous timescale is
given by \citep[p 88]{FKR02}:
\eqsl{eq:visc}{t_\mathrm{visc}&=&\nonumber
\frac{\sqrt{GMr}}{\alpha c_\mathrm{s}^2} \\
&=&  1.6\,\mathrm{Myr} 
\left(\frac{M}{60M_\odot}\right)^{1/2}
\left(\frac{c_\mathrm{s}}{1\, \mathrm{km\,s}^{-1}}\right)^{-2}
\left(\frac{\alpha}{0.1}\right)
\left(\frac{r}{0.1 \,\mathrm{pc}}\right)^{1/2}\, ,}
where a typical value for $\alpha$ is 0.1 and we have taken a radius
$r$ of order the average distance between massive stars.

Thus, the viscous evolution of the disc is happening on a similar
timescale as the equatorial mass ejection.
The re-formed discs near the massive stars are fed by both,
material \vsix{nuclearly} processed in the massive stars 
\vsix{and expelled} via
the equatorial mass ejections, and by pristine gas via accretion in
the shadowed equatorial region. It is thus plausible that
such discs develop a mixture of gas that 
has overall a very comparable contribution from
both, pristine and processed components, similar to
the composition of the second generation low-mass stars, which are presently observed
in GCs \citep[e.g.][]{Prantzea07}.

\subsection{Star formation in the discs}\label{sec:sfd}
Discs that are comparable in mass to the central object are known as
self-gravitating discs \citep[compare e.g. Sect. 3.1 in the review by][]{Armi11}.
The Toomre criterion \citep[e.g.][]{Shu92ii} readily shows that the inner 
discs are expected to be gravitationally unstable:
\eq{Q_\mathrm{T} = 0.7 
\left(\frac{c_\mathrm{s}}{10\, \mathrm{km\,s}^{-1}}\right)
\left(\frac{m}{10\, M_\odot}\right)^{-1}
\left(\frac{M}{60M_\odot}\right)^{1/2}
\left(\frac{r}{100R_\odot}\right)^{1/2}\, ,
}
where $m$ is the disc mass, which we assume to be equally distributed
within the scale radius $r$ for this estimate. We have assumed a
higher sound speed in the disc than calculated for the gas on larger
scales, because near the massive stars, the radiation will
heat the discs \citep[compare e.g.][]{AS12}. 
However, even for a sound speed of 10~km/s, the inner discs
reach the critical mass for gravitational instability on a timescale
of $10^6$~years. 

\subsubsection{Consequences of gravitational instability}
Gravitationally stable, as well as unstable discs have been studied in
the context of planet formation. The
latter case is however not yet fully understood 
\citep[compare e.g. the review by][]{KN12}. 
Spiral density waves with associated torques and strong
accretion are a likely outcome. It is therefore possible that some disc material is
accreted on to the FRMS, particularly in the ``off-phases'' of the
equatorial ejections. Yet, this material should then also have a high
residual angular momentum compared to the gas in the outer regions of
the star. Thus the time interval to the next ejection should be
shortened. However, the time interval between the ejections is already
much shorter than the viscous timescale (\rev{Eq.}~\ref{eq:visc})
and the timescale for the disc to become gravitationally
unstable. Hence, the disc sees an almost constant flux of material
from the star. It is thus possible that the main sink for the
disc mass would be the formation of gravitationally bound
objects. 

Likely, the disc will remain near $Q_\mathrm{T}\approx1$,
which implies a disc mass that declines with time, because also the
FRMS looses a large fraction of its mass. If the second
generation stars would form in this way, one would therefore expect
that the formation rate of second generation stars\vsev{, or
proto-stars which may continue to accrete for some time,}
is highest towards the beginning of the
equatorial ejections, i.e. about 2-3~Myrs after the birth of the first
generation of stars (1~Myr till the equatorial ejections start also
in the lower mass FRMS,
$\approx 1-2$~Myr viscous timescale and timescale for the disc to become
gravitational unstable). The equatorial ejections cease entirely
about \vsix{6}~Myr after the birth of the first generation of stars,
i.e. 2~Myrs into the SNe phase\vsev{, if massive stars explode (compare
Sect.~\ref{sec:SNe}).}
We show in Sect.~\ref{sec:tacc}
below that turbulence in this phase stops accretion on to the
discs. This effect also keeps the second generation stars
chemically clean from supernova
ejecta. Once
the discs are no longer fed, they are expected to form their
gravitationally bound objects and disappear within at most a few Myrs
\citep{HLL01}. The formation of the second generation stars is
therefore complete at, roughly, 10~Myrs after the birth of the first
generation of stars.

Thus, while we are not in a position to add all the details that still
need to be worked out in the context of the formation of self
gravitating bodies in discs, it appears viable that the
second generation of stars forms over a timescale of 6-10~Myrs in
discs around FRMS (Fig.~\ref{fig:sketch}, second row from top).

\subsubsection{The distribution of orbit eccentricities and the ratio
  of pristine-to-processed gas in the second generation stars}
\label{subsec:pri-to-pro}
\veight{We have seen in Sect.~\ref{subsec:accsup}, above, that the
  amount of gas a given FRMS disc accretes from the ICM depends strongly
  on the parameter $v_\mathrm{az,0}/v_\mathrm{circ}$. 
  The distribution function of $v_\mathrm{az,0}/v_\mathrm{circ}$ for
  the sample of FRMS in the cluster will therefore directly determine
  the distribution of the amounts of accreted ICM to the FRMS discs.}
  
  \veight{The actual abundance pattern in each second generation star will
  however not only be determined by the overall ratio of pristine gas 
  accreted from the ICM over ejected gas from the FRMS. The abundances
  in each disc will be also a function of time because of the time
  dependence of the abundances in the equatorial ejecta
  \citep{Decrea07a}. The
  composition of the equatorial ejecta is first very similar to the
  one of the pristine gas. As time goes on, the composition changes
  and shows increasing signatures of hydrogen burning. 
  \citet{Decrea07a} have shown that the observed spread in light
  element abundances may only be obtained if the time variation of the
  abundances in the ejecta is used. For example, only at late times do
  we yield oxygen depletions high enough to be compatible with the
  oxygen poor end of the abundance distributions in many
  \rev{GC}s \citep{GCB12}. 
If we would mix all the ejecta of a given star into a common
  reservoir, which would be the disc in our case, the average oxygen
  depletion would already be insufficient to explain many of the
  observed second generation stars. Further mixing with pristine gas
  would make the situation even worse.}
 \veight{We therefore require that the star formation in the FRMS
   discs happens sequentially. This means that for each second
   generation star, its formation is complete on a timescale
   significantly shorter than the lifetime of its parent FRMS disc, so
 that it may conserve the current abundance pattern of the disc at the
time of its formation. This may for example be achieved via migration
towards outer orbits, where the gas densities in the discs are low, or
encounters between second generation stars which might scatter them
out of the plane of the disc.}
 
\veight{A consequence of this way of star formation is that the
  distribution of abundances will always be continuous. Even in the
  extreme case when we assume a population of stars with circular
  orbits, which therefore do not accrete anything from the ICM,
  together with a population of stars with low eccentricities, which
  therefore accrete strongly, we would only expect that for the
  population which have more ICM accretion, the range of metallicities
would be reduced. It is difficult to imagine that this could lead to a
double peaked distribution like the one observed by \rev{\citet{Marinea08}}
in \object{M4}. We could however well explain the more uniform
distribution found in \object{NGC~1851} \citep{Lardea12}.
Details of abundance distributions will be the subject of future work.}

\section{Supernovae}\label{sec:SNe}
It is well established that massive stars ($M\gtrsim25M_\odot$)
  end their lives as black holes \citep[e.g.][]{Fry99}. \veight{Stars
  initially less massive than this are expected to produce energetic 
  SNe-events or, may be, Gamma ray bursts for fast rotators}
  \citep[e.g.][]{YLN06,DOCO12}. This is much less clear for
  the more massive stars \citep[e.g.][]{Fry99,Belcea12}, which might
  turn silently into black holes \citep[compare also the discussion
  in][and references therein]{Decrea10}.\\
  This introduces a considerable uncertainty into our model \CC{timeline}: The
  120~$M_\odot$ star would explode at 3.5~Myr, the 25~$M_\odot$ star
  at 8.8~Myr. The wind-bubble phase ends and the supernova phase begins at
  some point within this range with the first energetic supernova. In
  the following we will use the term supernova for simplicity, always
  subsuming the possibility of Gamma-ray bursts.

\veight{We will now first argue that the SNe will likely 
not significantly alter the picture
of the star formation in the FRMS discs (Sect.~\ref{sec:sn-sf}). The
SNe will cause substantial turbulence in the ICM
(Sect.~\ref{sec:sn_ICM}), which will likely lead to mixing of the SN-ejecta
with the cold phase of the ICM (Sect.~\ref{sec:sn-mix}). The SN-ejecta
may however not enter the second generation stars, because on the one
hand, the ICM does not form stars in this phase, and on the other
hand, the ICM can no longer accrete on to the FRMS discs because of
the large gas velocities (Sect.~\ref{sec:tacc}). From the end stages
in the lifes of the FRMS onwards, we expect the second generation
stars to be dispersed, first within the half-mass radius, and on the
relaxation timescale also throughout the whole cluster
(Sect.~\ref{sec:disp}).}

\subsection{The effect of supernovae on the associated 
  star-forming disc} \label{sec:sn-sf}
Supernova ejecta are fast and cannot be retained in the gravitational
potential of the discs around their parent stars
\citep[compare][]{Decrea07b,Decrea07a}.
Since the disc
occupies only a small solid angle, much of the energy will escape into the
other directions, especially if the energy release occurs in the form
of jets as might be the case for fast rotating massive stars 
\citep[e.g.][]{BKM08,TKS09,DOCO12},
and ablate only the surface of the disc. At this point (4.5~Myrs
for the $60M_\odot$ star) we might expect that a large fraction of the inner disc
gas has already accreted on to the newly formed stars. In this case,
some remaining debris might be cleared away by the explosion.
Yet, it is also possible that a substantial amount of gas is left. 
If there
would still be a substantial amount of gas in the disc, one would
expect, similar as in
the simulations of \citet{Gaiblea12} for galactic scales, 
a compression of the disc and stronger gravitational instability
during this phase.

%-------------------------------------------------------------
   \begin{figure}
   \centering
   \includegraphics[width=.5\textwidth]{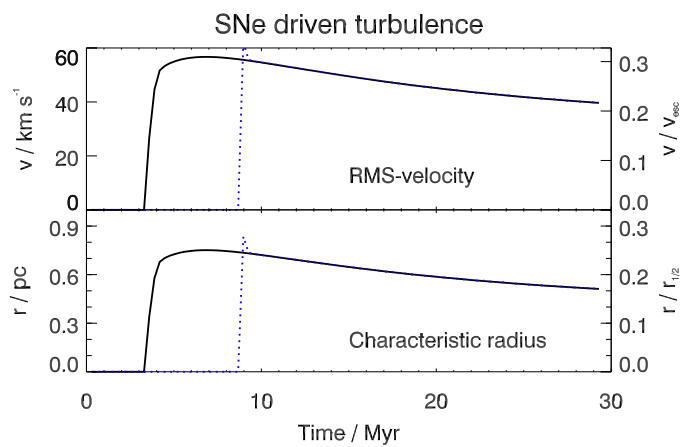}
      \caption{\vsix{RMS-velocity (upper part) and characteristic
        radius (lower part)
        established in the SNe driven convective phase. 
        The left axis shows, respectively, km/s and pc. The right axis
        indicates, respectively, velocity in units of the central
        escape velocity and radius in units of the half-mass radius.
        The solid black line is for the case when all massive stars
        produce energetic SNe. The dotted blue line assumes direct
        black hole formation without energy release to the ICM for all
        stars with $M>25M_\odot$. The plot shows that in the latter
        case, the onset of turbulence is delayed by 5.3~Myrs.}
        See Sect.~\ref{sec:sn_ICM} for more details.      
              }
         \label{fig:turb}
   \end{figure}
%-------------------------------------------------------------
\subsection{The effect of supernovae on the general ICM:
  turbulence}\label{sec:sn_ICM}
Supernovae are energetic enough to overcome the gravitational
potential of the cluster. They are expected to form a superbubble with
an expanding shell. As shown in \citetalias{Krausea12a}, such a shell would be
Rayleigh-Taylor unstable. The situation is similar to the one
described by the 2D-axisymmetric simulations of \citet{KPM01}: The fragments of the
shell are expected to fall back into the GC, and a convective cooling
core is expected to form. Now, the hot high entropy gas escapes, while
cooling shell fragments are continuously formed at the edges of
escaping bubbles, and fall back towards the centre, much like in a
boiling pot of water. 

We will now estimate, if the typical turbulent velocities could lead
to a significant increase of the scale height of the gas. 
We are particularly interested in the cold gas,
since most of the mass is expected to be in this phase.
Hydrodynamic energy transfer is expected between the
different gas phases \citep[e.g.][]{KA07}. In equilibrium, the energy input by
SNe should be balanced by radiative losses. But it is not
clear that such an equilibrium may be reached: In the simulations of
\citet{KPM01} the convection zone expands slowly, reflecting a steady
energy accumulation. \citet{KMM06}
summarise simulations of isothermal turbulence, and find a decay time
of $t_\mathrm{dis}=0.83 \lambda_{in}/v_\mathrm{rms}$. The injection
scale in our case is given by the half-mass radius, since this is
roughly the scale where the shells are destroyed by the
Rayleigh-Taylor instability. Following \citet{KMM06}, we set the
injection wavelength to $\lambda_\mathrm{in}=4r_{1/2}$, because
the gas is first in isotropic infall, and also because this is the maximum
possible wavelength. The rms-velocity is given by
$E=(1-\epsilon_\mathrm{sf}) M_\mathrm{tot} v_\mathrm{rms}^2/2$,
because the cold gas is supersonic so that the thermal energy may be
neglected. We may thus approximate the change of the energy in the gas
by 
\eq{\frac{\partial E}{\partial t}= \dot{E}_\mathrm{SN} -
  \frac{E}{t_\mathrm{dis}}\, .}
\vsix{Once the rms-velocity is calculated by this approach, 
the characteristic radius out to which a given gas filament may
ascend in the potential of the GC follows from:
\eq{\frac{r_\mathrm{rms}}{r_\mathrm{c}}
  =\frac{v_\mathrm{rms}}{\sqrt{{v_\mathrm{e,0}^2-v_\mathrm{rms}^2}}}\,
  ,}
where the central escape velocity $v_\mathrm{e,0}=183$~km/s for our
model GC.}
We plot the evolution of the rms-velocity and the characteristic
radius of the convective flow, for the case when the gas would be
 supported by turbulence
alone, in Fig.~\ref{fig:turb}.
Our choice of $\lambda_\mathrm{in}$
guarantees the smallest possible dissipation rate. The turbulent
velocities reached are large compared to normal interstellar medium (ISM)
turbulence, about 50~km/s. Yet, this falls short of the escape speed.
However, if the gas would have some other means of
support, apart from ram pressure due to turbulence, it would be quite
likely, that it would still be supported in this other
way. Alternative means of support against gravity could be via
magnetic fields or to some degree also radiation pressure \vsix{\citep{KT12}.}
\vsix{If this would be the case, our result might also} 
be interpreted in the way that SNe driven
turbulence is not able to increase the scale height of the gas in this
phase in any significant way.
\veight{We note that the characteristic turbulent velocity is mainly
  determined by the ratio of energy injection to gas mass, and also by
the size of the cluster via $t_\mathrm{dis}$. Thus, since GCs all have
a typical radius of a few pc, and since the ratio of energy injection
to gas mass depends only on the star formation efficiency, which
should be always around~1/3 for efficient ejection of the first
generation stars \citep{Decrea10}, we expect essentially always
turbulence at~50~km/s. For proto-cluster clouds with initial masses
above $10^6M_\odot$, which should be approximately the minimum initial
gas mass to form a \rev{GC}, this is below the escape velocity. Thus,
apart from very low-mass protocluster clouds with high star formation
efficiency, these findings should be generally applicable for all
\rev{GC}s.}

\subsection{Mixing of \CC{supernovae} ejecta with pristine gas}
\label{sec:sn-mix}
Turbulence is considered to be a decisive factor in mixing theories
\citep[for a review see][]{SE04}. Mixing of stellar ejecta into ICM
that later forms subsequent generations of stars is strongly
constrained by observations \citep[e.g.][and sect.~\ref{intro}, above]{GCB12}.
Here, we are interested in an
order of magnitude estimate and therefore use the mixing length
approach. We follow \citet{Xiea95} who apply mixing length
theory to interstellar clouds. Within this framework, the mixing
timescale is given by $\tau \approx H/v_\mathrm{d}$, where $H$
represents a mean of abundance and density scale heights, and
$v_\mathrm{d}$ is the diffusion velocity, given by $v_\mathrm{d}\approx
K/H$. The diffusion coefficient $K$ is simply approximated by the
product of characteristic turbulent velocity, $V_\mathrm{t}$ and the
mixing length $L$. Let us assume filament thicknesses of order 0.1~pc,
in the SNe driven turbulence phase, and set $L=H=$~0.1~pc. Thus, the
diffusion velocity equals the turbulent one. We have shown above that
the expected turbulent velocity in the SNe driven turbulence phase is
approximately 50~km/s, which therefore would be the characteristic
diffusion velocity in the cold gas. The sound speed in the hot
component will be larger than this. X-ray observations
\citep[e.g.][]{Jaskea11} generally find hot gas temperatures of order $10^6$~K
in superbubbles. The sound speed will consequently be several
100~km/s. Cold gas will therefore more efficiently be mixed into the
hot gas than vice versa.
 Using these assumptions, the mixing timescale is given by:
\eq{\tau \approx 2000 \, \mathrm{yrs} \, 
\left(\frac{H}{0.1\,\mathrm{pc}}\right)
\left(\frac{V_\mathrm{t}}{50\, \mathrm{km/s}}\right)^{-1} \, . }

This should be compared to the dynamical timescale, i.e. the time
needed by bubbles enriched with \CC{fast winds and SNe} ejecta to escape the GC. At
times when the GC is highly turbulent
the dynamical timescale will be given by:
\eqs{\tau_\mathrm{dyn,t}&=&r_{1/2}/V_\mathrm{t} \nonumber \\
&\approx& 60,000\, \mathrm{yrs} 
\left(\frac{r_{1/2}}{3 \, \mathrm{pc}}\right)
\left(\frac{V_\mathrm{t}}{50 \, \mathrm{km/s}}\right)^{-1}\, .}
When the turbulent velocities are small, the gravitational
acceleration on the cold gas which needs to refill the bubble volume
will be the limiting factor. Therefore, the Rayleigh-Taylor timescale
sets the minimum for the dynamical timescale:
\eqs{\tau_\mathrm{dyn,RT}&=&\sqrt{2 r_{1/2}/g} \nonumber \\
&\approx& 60,000\, \mathrm{yrs} 
\left(\frac{r_{1/2}}{3 \, \mathrm{pc}}\right)^{1/2}
\left(\frac{g}{5\times10^{-6} \, \mathrm{cm/s}^2}\right)^{-1/2}\, ,}
where we have scaled the gravitational acceleration $g$ to the average 
value within the half-mass radius for our model cluster.
Thus, the dynamical timescale is always around 60,000~yrs.

This means that in the supernova
phase, we expect the SNe ejecta to mix with the cold gas
phase of the ICM, because of fast turbulent mixing. On the other hand, in the absence of an efficient driver,
any turbulence should decay quickly and the gas velocities are
expected to be of order the sound speed, which is expected to be about
1~km/s (compare Sect.~\ref{sec:acc}). The mixing timescale increases
therefore to about 100,000~yrs, almost twice the dynamical timescale.
Hence, we expect no significant mixing in the wind phase before the
onset of the SNe.

We note that the issue of turbulent mixing in the ISM cannot be
regarded as settled yet. De Avillez \& Mac Low
(\citeyear{dAML02}) find in hydrodynamic
simulations of SNe-driven turbulence in the Milky Way disk diffusion
coefficients which are one to two orders of magnitude smaller than
in the mixing length approach. Correspondingly, the mixing timescales
we give above would underestimate the true timescales by one or even
two orders of magnitude. While an increase of one order of magnitude
would not change our result, two orders of magnitude would mean that
also in the SNe-driven turbulence phase, no significant mixing would
occur.

It has been suggested that stars with masses above $25 M_\odot$ might
not form SNe, but would instead collapse to a black hole without
noteworthy energy release (compare above).
%\citep[compare][ and references therein]{Decrea10}. 
This would delay
the onset of strong turbulence until about 8.8~Myr after the birth of
the first generation stars, the time when the $25 M_\odot$ stars would
explode. The FRMS could in this case accrete from the ICM up to this
point. Yet, the lack of turbulence would also prevent mixing, so that
the stellar ejecta carried by the fast radiative winds of the more
massive stars at this time, which carry He-burning products that are
not found in the second generation stars, cannot enter the FRMS-discs.
The same argument applies for the end of the wind bubble phase for the
case when the high mass FRMS do develop SNe.

\subsection{The effect of turbulence on accretion}\label{sec:tacc}
As discussed in 
Sect.~\ref{sec:acc}, accretion is significantly suppressed, if the
relative velocities between the accreting objects and the accreted gas
are large. This is the case for the SNe phase, as already the typical
gas velocities are around 50 km/s. Thus, accretion on to
the FRMS discs may only happen during the first few Myr 
\vsix{(4 to 9 ~Myr, depending on the mass limit for energetic SNe events)}
after birth of
the first generation of stars, before the
supernova phase. The associated smaller amount of accreted gas may
however be compensated by a slight increase of the orbital
eccentricities of the FRMS (Sect.~\ref{sec:acc}).
% This places the mass of the accretion discs around
% the FRMS in the range of tens of solar masses, with similar
% contributions from accreted ICM and stellar ejecta.
% The discs have therefore about 30~Myrs of basically undisturbed
% evolution. Yet, it is well known that lifetimes of circumstellar discs around stars
% of all masses are much shorter, limited by about 6~Myrs \citep{HLL01}.
% Hence, the discs should have disappeared, and have formed their stars,
% long before the end of the supernova phase.
We have shown in Sect.~\ref{sec:sfd} that the FRMS discs have
disappeared long before the end of the SNe phase. Once \vsev{a given}
FRMS \vsev{has} 
exploded as a supernova, there is therefore at most a short period of activity,
and the dark remnants then remain in-active until the \vsev{last SN has exploded}.

\subsection{Dispersal of the second generation stars}\label{sec:disp}
The FRMS loose about half of their mass via equatorial ejections 
\vsix{loaded with hydrogen-burning products} \CC{during the main sequence and the luminous blue
variable stage}, and this is the main mass loss channel in this \CC{early} phase. We expect that
towards its end, each FRMS disc has about the same mass as the FRMS,
perhaps even a bit more. A considerable fraction of the disc mass should
already have been used up to form the second generation stars. Thus,
even towards the end of the equatorial ejections, it is expected
that gravitational interactions lead to migration of the second
generation stars in the discs, and quite possibly also to \vsix{some 
ejections from the discs}. 

From the end of the equatorial ejection phase, the FRMS go on loosing
mass \vsix{via fast radiative winds and perhaps eventual SNe ejections}. 
The final dark remnants \vsix{might then} have only about ten per cent of the
mass that the FRMS had at the end of the equatorial ejection
phase. Consequently, the gravitational binding energy, which is to
first order simply
proportional to the square of the total mass, \vsix{would have} decreased by
a factor of a few. Virial equilibrium at the end of the equatorial
ejection phase demands that the kinetic energy is half of the binding
energy. Therefore, \vsix{after a SN-ejection}, 
the total energy of the system \vsix{would} be positive. 
\vsix{Even without SN-ejection, the mass loss of the FRMS after
the equatorial ejections have ceased is considerable. Consequently,
the second generation star orbits widen. This increases the cross
sections for collisions. Most of the
second generation stars may be 
lost from the parent star, but would remain centrally concentrated
within the half-mass radius, as also the FRMS and the dark
remnants (compare Fig.~\ref{fig:sketch}). 
The mixing of the first and second generation
stars due to two-body relaxation effects has been studied in detail by
\citet{Decrea08}. They show that the two stellar populations need two
relaxation timescales to mix, which corresponds to about 2.8~Gyr, from
their \rev{Eq.}~(1).
Therefore, there is not enough time for the second generation stars to
spread over the cluster, before the majority of the first generation stars are lost
from the cluster during the gas expulsion phase (section~\ref{sec:dmacc}).}
Yet it appears also well possible that the dark remnants retain some
of the second generation stars, which might thus form X-ray binaries
later. X-ray binaries are well known in GCs 
\citep[e. g.][]{BGM12,Barret12,Stacea12}.

\section{Accretion on the dark remnants}\label{sec:dmacc}
After the last type~II SNe has exploded, roughly 35~Myr after
the formation of the first generation of stars, turbulence
decays. With the formula given in Sect.~\ref{sec:sn_ICM} above, the
decay time evaluates to 0.2~Myr/$v_{50}$, where $v_{50}$ is the
rms-velocity in units of 50~km/s. 

In principle, black holes and
neutron stars may receive natal kicks as large as 
$\approx 400$~km~s$^{-1}$ \citep{RDS12}, large enough for them to
escape from any given GC. However, many neutron stars are actually
observed in GCs \citep[e. g.][]{BGM12,Barret12,Stacea12}. The recent
detection of two black holes in the Milky Way \rev{GC} M~22 
\citep{Stradea12b} has
called into question if all black holes receive large natal
kicks. Especially if black holes form directly without SN-ejection, no
significant natal kicks are expected. 

In the following, we assume
therefore that the dark remnants have the same
orbits as the FRMS have had before. Therefore, accretion should
again set in. 
Assuming 1.4~$M_\odot$ for the neutron stars, and 3~$M_\odot$ for the
stellar mass black holes and accretion over the full solid angle,
\rev{Eq.}~\ref{eq:Bondi} predicts about 10-100~$M_\odot$/Myr. If we
assume a similar suppression as in Sect.~\ref{sec:acc}, above, 
we end up with 0..003-0.03~$M_\odot$/Myr.
% or $\approx 0.01-0.1 M_\odot$ that each star receives when it traverses
%once the outer part of its orbit. 
The dark remnants will start their energy output after
about a viscous timescale, which is needed by the gas to get close
enough to their surface, or respectively, the horizon.  
The viscous timescale is about 1~Myr, according
to \rev{Eq.}~\ref{eq:visc}. Therefore, each dark remnant will receive 
0..003-0.03~$M_\odot$ of gas and then get activated.
This compares to the Eddington
accretion rate of $0.02 M_\mathrm{DR}
M_\odot$/Myr, where $M_\mathrm{DR}$ is the mass of the dark remnant in
solar masses. 
This means that once loaded with gas, the dark remnants \rev{could} be
active at the Eddington rate for about 0.1-0.5~Myrs
(Fig.~\ref{fig:sketch}, bottom).
We have shown in \citetalias{Krausea12a} that such a sudden onset of
accretion on the stellar mass black holes and neutron stars may
plausibly release a sufficient amount of energy to eject the ICM in at most
0.06~Myrs, corresponding to about half a crossing time. The
mass accreted on to the dark remnants may thus plausibly be sufficient
for this to occur.
The expulsion timescale is short enough to remove also the outer first
generation low-mass stars, which are not to appear in the GCs today.
We have also shown that a gradual activation of the dark
remnants as they become available, reinforcing the SNe in that phase,
is still insufficient to overcome the Rayleigh-Taylor instability.
For the efficiency of energy transfer we have adopted in \citetalias{Krausea12a} (20
per cent of the Eddington luminosity), the
limiting proto-cluster cloud mass up to which the gas expulsion worked
was about $2\times10^7M_\odot$. This would mean that gas expulsion
would work for most of the observed clusters, apart from the most
massive ones, e.g. \object{$\omega$ Centauri} \vsix{and
\object{M~22}}, where also a spread in
iron abundances is observed \vsix{\citep[and references
  therein]{GCB12}}. \rev{The repeated star-bursts expected in this case are
similar to the predictions for nuclear star clusters without
supermassive black holes by \citet{LKT82}.}

\vsix{\rev{For GCs below this mass limit, f}urther gas accretion events are expected from about 40~Myrs
  after the birth of the first generation stars due to the slow
  AGB-winds \citep{DErcolea08}. Any such events would re-activate the
  dark-remnants. Both, the accretion rate and the binding energy are
  \rev{approximately} linear in the gas density\rev{, if the stellar
    potential dominates. Otherwise, the binding energy drops even
    stronger with gas density. T}herefore, we expect
  again quick gas expulsion. This would be similar to the case of
  super-massive black holes in nuclear star clusters and elliptical
  galaxies, which may also
  be activated by AGB-winds
  \citep[e.g.][]{GCK05,Davea07,Schartea10,Daviesea10}. 
  Black-hole feedback may suppress star formation
\citep[e.g.][]{DSH2005}.}

\section{Discussion}\label{sec:disc}

\rev{\subsection{Orbit eccentricities of the FRMS}}
The weakest point of the presented scenario is perhaps the fine-tuning
required for the orbital eccentricities of the FRMS, in order to
obtain high enough accretion rates, 
\vsix{both, on to the FRMS discs and also the dark remnants}
(Sect.~\ref{sec:acc}). We require
velocities near the outer turning points of less then about a tenth of the circular
velocity there. The former evaluate to a few km/s.
An important point to make in this context is that the requirements are
the very same as for the dark remnant accretion, as the FRMS turn
into the dark remnants at the end of their lives. If gas expulsion is
indeed due to the dark remnants, as we have proposed in \citetalias{Krausea12a}, then
the orbits are also ok for accretion on to the FRMS, unless they were
significantly affected by the transformation (compare sect.~\ref{sec:dmacc} above).

\rev{\subsection{Support of the gas and relation to GC formation 
scenarios}}
The FRMS orbits will of course be related to the formation of the
GC. This is so far unclear, and different scenarios are
considered \rev{\citep[compare e.g.][]{HHM98,mypap02a,KG05,Gnedin11,GS11,HH11}}.
In general, one may consider two classes of physical conditions:
First, the first generation stars might be formed while the gas is in
free collapse. \vsev{This is among the scenarios currently discussed
for star formation in general \citep[e.g.][]{ZAVC12}.}
Then, by definition, the stars will be born with large
radial velocities, and much smaller azimuthal velocities, roughly as required.
The second option is to have internal support against gravity in the
protocluster cloud. As the decay time for turbulence is short 
(compare Sect.~\ref{sec:sn_ICM}) and an unusually high level of
turbulence is required ($\approx 100$~km/s), it seems unlikely that
such a protocluster would be supported by turbulence. If it was, the
stars would be expected to form with comparable radial and azimuthal
velocities, which would be a serious problem for this scenario.
Thermal pressure seems also unlikely, as the cooling times would be
very short.
The protocluster cloud might also be
supported by magnetic fields. In general
ISM simulations, the coldest and hence densest phase is found to be 
magnetically supported, in contrast to turbulent support for intermediate
temperatures \citep{dAB05}. \vsev{This agrees with observations of
  molecular clouds \citep{Crutcher12}. In a magnetically supported protocluster
cloud, the stars would lose their magnetic support immediately after
formation, and one would also expect highly eccentric orbits.}

\rev{\subsection{Effects of magnetic fields}}
\vsev{Magnetic support would mean that the proto-cluster cloud could keep
the same scale height as the stars until the gas is expelled from the
GC. This would be highly beneficial, as for our scenario accretion
takes place mainly when the stars are in the vicinity of the core
radius.
Magnetic fields might however also inhibit accretion. \citet{Cunea12} show
that Bondi accretion from an isothermal, magnetised gas is suppressed
by a factor $0.4 c_\mathrm{s}/c_\mathrm{A}$, where $c_\mathrm{A}$
refers to the Alfv\'en speed. This might seem moderate compared to the
$ (c_\mathrm{s}/v)^3$-factor for bulk motions at velocity $v$. Yet, to
support the protocluster cloud magnetically, this also amounts to two
orders of magnitudes, which would significantly restrict the available
orbit eccentricities. Ambipolar diffusion should however be important
in this context, as the majority of the cold gas would be neutral. 
Also, \citet{Cunea12} assume that the magnetic flux
accumulates around the accretor, which is the underlying physical
process inhibiting the accretion. In our case, we would expect the
flux to be advected into the accretion discs, there to be dissipated
by turbulence.  Qualitatively, we therefore believe that a
magnetically supported protocluster cloud would be a viable
alternative in the present context.
Radiation pressure might also help to
support the gas after the first generation massive stars have formed \citep{KT12}.}

\rev{\subsection{Stronger mass segregation as alternative to 
supporting the
gas on a scale of the half-mass radius}\label{sec:mseg}
An alternative to supporting the gas on the scale of the
  half-mass radius would be to assume a stronger concentration of the
  gas as well as the massive stars towards the centre of the GC,
  implying stronger mass segregation: 
In our model setup (Sect.~\ref{sec:dmacc}), we specified
that the massive stars are confined to the sphere inside the half-mass
radius. In fact, \citet{Leighea13} in their model for  dark-remnant accretion
and dynamical black-hole ejection in GCs, assume a much
stronger mass segregation. This might happen, if a GC is formed from merging
sub-clumps. In this case, strong mass segregation of the stars is expected to arise in
about $10^6$~yr \citep{Allisea10,Girichea12,Pangea13_pre}. Assuming a
velocity dispersion inversely proportional to the stellar mass, as in
\cite{Leighea13}, would result in our relevant FRMS population having
maximum velocities of order the sound speed. In such a scenario, both,
the massive stars and the gas could be very much concentrated towards
the cluster centre. Hence, accretion would no longer be affected by
the orbital parameters, as the FRMS velocities would always be small.
Apart from the stronger accretion, the formation of the second-generation stars 
would in this variant of the scenario proceed
in almost the same way as in the standard case
(Sects.~\ref{sec:discmerge} and~\ref{sec:sfd}). The star forming discs
must have some way to prevent complete mixing of the ejected and the
accreted gas in order to explain the most oxygen poor stars (compare
Sect.~\ref{subsec:pri-to-pro}). The SN-phase would be essentially
unaffected, whereas the accretion on to the dark remnants would also
be enhanced. The latter might lead to an observationally interesting
collective active phase of order the viscous timescale, i.e. about 1~Myr.}

\rev{
\subsection{Limitations of dark-remnant accretion due to local
  radiative feedback}\label{sec:dmacclim}
Radiative feedback has been found to limit accretion on to
  stellar-mass black holes in spherical and clumpy accretion flows
  \citep{Milosea09}. This applies to the dark-remnant accretion phase,
as we have assumed spherical accretion in
Sect.~\ref{sec:dmacc}. \citet{Milosea09} show that the average
accretion rate may be limited by a value
two or three orders of magnitude below the
Eddington limit. They caution however that angular momentum,
turbulence, and thermodynamic phase segregation might affect the
result. They also find that on short timescales the accretion
rate may exceed their limit substantially. We require only a very
small active phase of the dark remnants
to expel the gas ($<10^5$~yr). Further work is clearly required to
settle the case. The strong
observational constraints on the present-day gas content of GCs and the
Fe-uniformity of the present-day stars put strong constraints on the
gas expulsion mechanism (compare \citetalias{Krausea12a}).
\subsection{Top-heavy IMF}\label{sec:th-IMF}
Gas expulsion could be substantially eased by a non-standard,
top-heavy IMF.
For the model presented in this paper (Sect.~\ref{sec:dmacc}), we have
used a normal IMF, following other recent work 
\citep[e.g.][]{Decrea10,DErcolea11}. 
However, the first generation low-mass stars are of no importance for our
scenario. In fact, there is even an issue with their ejection, which
we have found here not to work with
SN-feedback alone (compare paper~I and Sect.~\ref{sec:dmacc}, above), 
and which may therefore require the coherent activity of the dark 
remnants. The ejected low-mass stars might
contribute significantly to the stellar population of the Galactic
halo \citep{SC11}, but there is no strong constraint on how many stars
have to escape.
The IMF for the first generation
stars might therefore have been top-heavy \citep[e.g.][and references
in Sect.~\ref{intro}]{PC06}. 
This would not necessarily be in contradiction with
recent work on clusters formed under typical present-day Milky Way
conditions \citep{Hen12,KKMcK12}, as the conditions in which the
Galactic-halo GCs formed are not known. The central disc in the Milky
Way's nuclear star cluster, a massive and concentrated star cluster
similar in this respect to GCs, is an example where observational evidence
for a top-heavy IMF exists \citep{Bartkea10}. A high star-formation
efficiency, as often assumed for GCs, is also discussed for this
region \citep[e.g.][]{Silkea12}. A top-heavy IMF has been suggested in
the context of GCs on the basis of the energy requirements for the gas
expulsion \citep{Marksea12}.\\
In the extreme case, assuming that almost no first generation low mass
stars form, the initial mass of our model-protocluster cloud ($9
\times10^6 M_\odot$, Sect.~\ref{sec:setup}) could be
reduced by one order of magnitude. Because of the combined effect of
lower gas densities and smaller stellar velocities,
the accretion suppression factor would be constrained to be larger by
a factor of a few, moving the vertical line in
Fig.~\ref{fig:path} to the right. This would narrow the allowed
orbital parameters, requiring somewhat higher eccentricities 
(but see Sect~\ref{sec:mseg}
for a non-exclusive model variant which has the opposite effect). 
Depending on the star-formation efficiency and the
energy-injection efficiency in the SN-phase, the SNe might already eject
the gas. We stress that this refers to the extreme scenario. We have
shown in paper~I that even tripling the massive stars would not yet
suffice to expel the gas via supernova feedback. 
The argument is still valid that favourable conditions for
accretion on to the FRMS imply later accretion on to the dark
remnants. 
A possibly reduced gas density due to successful SNe-feedback would
cause less accretion on to the dark remnants (linear dependency,
compare Eq.~(\ref{eq:Bondi})), but the binding energy would also be
reduced. Hence, the dark-remnant feedback would be expected to keep
the GC gas-free.}

\rev{\subsection{Comparison to Be-star decretion discs and
    ``first-star'' simulations}}
A key point in our scenario is the formation of the second
generation stars in discs, fed by both, accretion of pristine gas and
equatorial ejections of the FRMS. Such equatorial ejections are known from
Be-type stars \citep[e.g.][]{Riviea01}. Recently, the spreading out of
such discs is also attempted to be understood in terms viscous evolution
\citep{Haubea12}, similarly as we propose here for the FRMS discs.\\
The formation of stars as opposed to planets 
in circumstellar discs might seem unusual. Yet, detailed models have
recently be computed in the context of the formation of the first
massive stars \citep[e.g.][]{Greifea12}\rev{, which also may have been
  very fast rotators \citep{SBL11}}. For example,
\citet{SGB12} simulate
the formation of a $\approx30 M_\odot$ star with radiative
feedback. They find that the radiative feedback does not stop
the secondary star formation in the \vsix{accretion} disc. During their simulation time
of a few 1000~yrs, they form a few proto-stars of sometimes several solar
masses. \citet{Clarkea11} form four secondary proto-stars within 110~yrs in
a similar simulation. Some of these proto-stars may end up in
the primary star. Adapted to the FRMS case, this might then lead to a
more rapid sequence of ejections. Clearly, a lot of details need to be
worked out, some are not comparable (e.g. the first stars are usually
assumed to form in dark matter halos). Yet, we believe that these
developments are encouraging.

\section{Conclusions}\label{sec:conc}
Using FRMS data, we have developed a comprehensive scenario for the
formation of second generation \CC{GC} stars (compare Fig.~\ref{fig:sketch}):
\begin{itemize}

\item The ICM obtains a spongy structure
     in the {\em wind phase} (Fig.~\ref{fig:sketch}, second row)
     due to the space-filling wind bubbles of the
     massive stars. 
\item We confirm the analysis of Conroy and Spergel that
     the Lyman--Werner photon flux density is sufficiently high that
\rev{most of the}
     hydrogen molecules dissociate and no stars may form in the
     normal way, during \veight{the wind and the supernova phase}. 
\item If the \vsix{first generation massive} stars have however 
     equatorial ejections \vsix{as expected in the case of fast
       rotation}, we show that accretion on to their discs 
     \rev{resumes} in the shadow of the equatorial ejecta. Massive
     discs are formed fed by both, FRMS ejecta from the inside, and
     accretion of pristine gas from the outside. Within an order of
     magnitude estimate, both contributions can be comparable, as
     required by self-enrichment calculations \citep{Decrea07a}, if
     the orbits of the FRMS are sufficiently eccentric.
\item The second generation stars may then
     form due to gravitational instability in these discs.
     (Fig.~\ref{fig:sketch}, second row). This
     second-generation
     star formation might carry on for a few Myr through the 
     {\em supernova phase}, \veight{because SNe ejecta are not expected to
     enter the discs. The increased pressure due to the SN may however
     also compress its associated disc such that the last stars in the
     respective disc form at this occasion.}
\item \veight{The second generation stars form sequentially in their
    parent FRMS discs, with the current light element abundances of
    the respective discs. We propose this as the physical mechanism
    underlying the detailed abundance calculations of
    \citet{Decrea07a}.}
\item \veight{The formation of the second generation stars is
    completed latest about 10~Myr after the formation of the first
    generation. The second generation stars are first dispersed within
  the half-mass radius, and much later, after the gas expulsion, also
  throughout the entire GC.}
\item The SNe drive turbulence at about 50~km/s 
  \rev{(Fig.~\ref{fig:sketch}, third row)}. The remaining gas outside
     the discs might then mix with the SN ejecta due to
     turbulence caused by the SN energy. The gas
     remains bound to the cluster until the SNe have ceased,
     turbulence has decayed and the gas can once more accrete suddenly
     on to the dark remnants. The gas may not accrete on to the FRMS
     discs before due to the high turbulent velocities. In this way,
     the second generation stars are prevented from pollution by SN
     ejecta and He-burning products in general. 
     \veight{These findings should hold for
       all globular clusters, apart from perhaps the ones at the low
       mass end if their star formation efficiency significantly
       exceeds~1/3.}
\item  In the {\em dark remnant accretion
       phase}
      (Fig.~\ref{fig:sketch}, bottom), the gas
     is efficiently expelled due to the strong energy release
     associated with the accretion on to the dark remnants. This
     happens fast enough so that a large fraction of less tightly
     bound first generation stars \rev{are} also lost.
 \end{itemize}
% The case that the SNe fail to eject the gas, but a later
%      accretion event succeeds would pose no difficulty for the FRMS
%      scenario, as the formation of the second generation stars is
%      taking place in circumstellar discs set up well before the
%      supernova phase. These discs are insulated from the ICM in the
%      supernova phase, because the large turbulent velocity inhibits accretion. 

% The AGB scenario would consequently have to put
%      the formation of the second generation stars after the dark
%      remnant phase. Yet, if the ICM has once preferred to accrete on to
%      the dark remnants, followed by quick gas expulsion,
%      why should it not do the same again in such a
%      cooling flow phase? If so, fewer stars might be expected to
%      form in this phase and the gas inflows should trigger repeated
%      short outbursts.

Potential problems for the scenario include:
\begin{itemize}
\item For efficient accretion, the orbits of the massive stars need
  to be eccentric, with
  low angular momenta, and velocities near the outer turning points
  below about 10~\% of the circular velocity. This may relate to the
  formation scenario. It would pose a serious difficulty for this
  model, if such low angular momenta could be ruled out
  observationally \rev{and, consequently, stronger mass segregation would have to be
    assumed (Sect.~\ref{sec:mseg})}.
\item Further, the gas scale height has to be comparable to the core
  radius, as the main gas accretion would occur near the core
  radius. The necessary support of the gas against gravity may be due
  to radiation pressure \citep{KT12} and magnetic fields, details are
  however beyond the scope of this work. \rev{Stronger mass
    segregation would again alleviate this problem}.
\item Star formation in discs around massive stars is only recently
  being explored. While the results from the ``first stars''
  simulations  are
  encouraging, simulations tuned to the specific conditions in FRMS
  are not yet available.
\item   If for some reason gas expulsion by dark remnant accretion would
   not work \rev{(Sect.~\ref{sec:dmacclim})}, all gas should at some point form stars from the
   quite possibly 
   SN enriched gas \rev{(Sect~\ref{sec:sn-mix} for mixing uncertainties)}, unless some other way of gas expulsion
   would be found. This would however conflict with observations
   except for the \CC{rare and } most massive GCs \CC{like $\omega$~Centauri and M22 that exhibit Fe spread}. 
\end{itemize}
% Multiple
%    star-formation episodes, each one clomplete with a classical first
%    generation and second generation of stars, as infered e.g. for the
%    most massive known GC
%    \object{$\omega$~Centauri} \citep{Marinea12}, 
%    seem to be more easily explained in the
%    context of the FRMS scenario, though it remains to be seen if the
%    observed abundance tracks may be explained by a detailed mixing model.

\begin{acknowledgements}
We thank the referee, Volker Bromm, for very useful comments which
helped to improved the manuscript.
Also, we gratefully acknowledge discussions with
Mordecai-Mark Mac Low.
 This research was supported by the cluster of excellence ``Origin and
Structure of the Universe'' (www.universe-cluster.de) and  the ESF
EUROCORES Programme 'Origin of the Elements and Nuclear History of the
Universe' (grants 189 and 190).
CC and TD also acknowledge financial support from the Swiss 
National Science Foundation (FNS) and the French Programme 
National de Physique Stellaire (PNPS) of CNRS/INSU.
\end{acknowledgements}

\bibliographystyle{aa}
\bibliography{/Users/mkrause/texinput/references}
%\bibliography{references}

% \appendix

% \section{The accretion suppression factor for elliptical orbits}

% The Bondi accretion rate (compare \Rev{Eq.}~\ref{eq:Bondi}) is reduced
% by a factor $(v/c_\mathrm{s})^{-3}$ for a relative velocity $v$
% between the accretor and the gas. Here we estimate how the suppresion
% depends on the eccentricity, averaged over one orbit.
% In a spherical potential, the orbits are generally
% expected to be of the rosette type \citep{BT08}, with most of the
% stars having only slightly negative total energies.
% Approximating the orbit by an ellipse, with the radius $r$ as a
% function of azimuth $\phi$ given by $r=p
\end{document}

%% file: def.tex
\bibpunct{(}{)}{;}{a}{}{,}
\newcommand{\ignore}[1]{}
\newcommand{\eq}[1]{\begin{equation} #1 \end{equation}}
\newcommand{\eql}[2]{\begin{equation} \label{#1} #2 \end{equation}}

\newcommand{\eqs}[1]{\begin{eqnarray} #1 \end{eqnarray}}
\newcommand{\eqsl}[2]{\begin{eqnarray} \label{#1}#2 \end{eqnarray}}

\newcommand{\dm}[1]{\begin{displaymath} #1 \end{displaymath}}
\newcommand{\sqrbrk}[1]{$\left[\right.$#1$\left.\right]$}